\documentclass[12pt,a4paper]{article}
\usepackage{epsfig,graphics,amsmath,amssymb,here}

\makeatletter
\newif\if@preliminary
\@preliminaryfalse
\def\preliminary{\@preliminaryfalse}
\def\preprintno#1{\def\@preprintno{#1}}
\def\address#1{\def\@address{#1}}
\def\email#1#2{\thanks{\tt #1@{}#2}}
\def\abstract#1{\def\@abstract{#1}}
\renewcommand\abstractname{ABSTRACT}
\newlength\preprintnoskip
\newlength\abstractwidth
\@titlepagetrue
\renewcommand\maketitle{\begin{titlepage}%
  \let\footnotesize\small
  \hfill\parbox{\preprintnoskip}{%
  \begin{flushright}\@preprintno\end{flushright}}\hspace*{1cm}
  \vskip 60\p@
  \begin{center}%
    {\Large\bf\boldmath \@title \par}\vskip 1cm%
    {\sc\@author \par}\vskip 3mm%
    {\@address \par}%
    \if@preliminary
      \vskip 2cm {\large\sf PRELIMINARY DRAFT \par \@date}%
    \fi
  \end{center}\par
  \@thanks
  \vfill
  \begin{center}%
    \parbox{\abstractwidth}{\centerline{\abstractname}%
    \vskip 3mm%
    \@abstract}
  \end{center}
  \end{titlepage}%
  \setcounter{footnote}{0}%
  \let\thanks\relax\let\maketitle\relax
  \gdef\@thanks{}\gdef\@author{}\gdef\@address{}%
  \gdef\@title{}\gdef\@abstract{}\gdef\@preprintno{}
}%
\def\@citex[#1]#2{\if@filesw\immediate\write\@auxout{\string\citation{#2}}\fi
  \def\@citea{}\@cite{\@for\@citeb:=#2\do
    {\@citea\def\@citea{,\penalty\@m}\@ifundefined
       {b@\@citeb}{{\bf ?}\@warning
       {Citation `\@citeb' on page \thepage \space undefined}}%
\hbox{\csname b@\@citeb\endcsname}}}{#1}}
\def\citerange{\@ifnextchar [{\@tempswatrue\@citexr}{\@tempswafalse\@citexr[]}}
\def\@citexr[#1]#2{\if@filesw\immediate\write\@auxout{\string\citation{#2}}\fi
  \def\@citea{}\@cite{\@for\@citeb:=#2\do
    {\@citea\def\@citea{--\penalty\@m}\@ifundefined
       {b@\@citeb}{{\bf ?}\@warning
       {Citation `\@citeb' on page \thepage \space undefined}}%
\hbox{\csname b@\@citeb\endcsname}}}{#1}}
\long\def\@makecaption#1#2{%
  \vskip\abovecaptionskip
  \sbox\@tempboxa{#1: \emph{#2}}%
  \ifdim \wd\@tempboxa >\hsize
    #1: \emph{#2}\par
  \else
    \hbox to\hsize{\hfil\box\@tempboxa\hfil}%
  \fi
  \vskip\belowcaptionskip}
\def\fmslash{\@ifnextchar[{\fmsl@sh}{\fmsl@sh[0mu]}}
\def\fmsl@sh[#1]#2{%
  \mathchoice
    {\@fmsl@sh\displaystyle{#1}{#2}}%
    {\@fmsl@sh\textstyle{#1}{#2}}%
    {\@fmsl@sh\scriptstyle{#1}{#2}}%
    {\@fmsl@sh\scriptscriptstyle{#1}{#2}}}
\def\@fmsl@sh#1#2#3{\m@th\ooalign{$\hfil#1\mkern#2/\hfil$\crcr$#1#3$}}
\makeatother

\def\fmfL(#1,#2,#3)#4{\put(#1,#2){\makebox(0,0)[#3]{#4}}}

\newcommand\OBS{\mbox{$\cal{O}$}}
\newcommand\PRO{\mbox{$\wp$}}
\newcommand\ZJ{\mbox{Z $\rightarrow$ 4 jets}}
\newcommand\ZG{\mbox{$Z \rightarrow b \bar{b} G G$}}
\newcommand\ZB{\mbox{$Z \rightarrow b \bar{b} b \bar{b}$}}

\newcommand\ZU{\mbox{$Z \rightarrow b \bar{b} u \bar{u}$}}
\newcommand\ZD{\mbox{$Z \rightarrow b \bar{b} d \bar{d}$}}

\newcommand\EZG{\mbox{$e^+ e^- \rightarrow Z \rightarrow b \bar{b} G G$}}
\newcommand\EZB{\mbox{$e^+ e^- \rightarrow Z \rightarrow b \bar{b} b \bar{b}$}}
\newcommand\EZQ{\mbox{$e^+ e^- \rightarrow Z \rightarrow b \bar{b} q \bar{q}$}}

\newcommand\bb{\mbox{$\bar{b}$}}
\newcommand\bq{\mbox{$\bar{q}$}}
\newcommand\ycut{\mbox{$y_{cut}$}}
\newcommand\hvb{\mbox{$\hat{h}_{Vb}$}}
\newcommand\hab{\mbox{$\hat{h}_{Ab}$}}
\newcommand\hb{\mbox{$\hat{h}_{b}$}}
\newcommand\hbn{\mbox{$\tilde{h}_{b}$}}
\newcommand\dhb{\mbox{$\delta\hat{h}_{b}$}}
\newcommand\dhbn{\mbox{$\delta\tilde{h}_{b}$}}
\def\ltap{\raisebox{-.4ex}{\rlap{$\sim$}} \raisebox{.4ex}{$<$}}

\begin{document}
%\shortletter        % subdivided in paragraphs instead of sections
\preliminary        % mark on title page
\baselineskip20pt   % stretch linespacing in main text
%%%%%%%%%%%%%%%%%%%%%%%%%%%%%%%%%%%%%%%%%%%%%%%%%%%%%%%%%%%%%%%%%%%%%%%%
\preprintno{HD--THEP 98--63\\hep-ph/9901343\\[0.5\baselineskip] December 1998}
\title{%
 CP VIOLATION IN DECAYS Z $\rightarrow$ 4 JETS 
\footnote{Supported by German Bundesministerium f\"ur Bildung und
Forschung (BMBF),\\ Contract Nr.~05~7HD~91~P(0), and by the
Landesgraduiertenf\"orderung}}
\author{%
 O.~Nachtmann\email{{\rm Email: }O.Nachtmann}{thphys.uni-heidelberg.de}
 and C.~Schwanenberger\email{{\rm Email: }C.Schwanenberger}{thphys.uni-heidelberg.de}
}
\address{%
 Institut f\"ur Theoretische Physik, Universit\"at Heidelberg,
 Philosophenweg 16\\
 D--69120 Heidelberg, Germany
}
\abstract{%
We analyse CP-violating effects in Z $\rightarrow$ 4 jet 
decays, assuming the presence of CP-violating effective 
$Z b \bar{b} G$ and $Z b \bar{b} G G$ couplings. We discuss
the influence of these couplings on the decay width.
Furthermore, we propose various strategies of a direct 
search for such CP-violating couplings by using different
CP-odd observables. The present data of LEP 1 should give
significant information on the couplings. 
%Finally, an 
%analysis of a CP-violating triple gluon coupling is 
%presented.
%
%We analyse CP-violating effects in Z $\rightarrow$ 4 jet decays, assuming 
%the presence of CP-violating effective $Z b \bar{b} G$ and $Z b \bar{b} G G$
%couplings. We discuss the influence of these couplings on the decay width.
%Furthermore, we investigate various CP-odd observables and compare their
%sensitivities to optimal observables. Finally, we propose the measurement of
%CP-violating axial vector and vector couplings with a CP-odd tensor and
%a CP-odd vector observable; this measurement should be possible with the
%present data of LEP 1.
}
\maketitle

%%%%%%%%%%%%%%%%%%%%%%%%%%%%%%%%%%%%%%%%%%%%%%%%%%%%%%%%%%%%%%%%%%%%%%%%
%%% Text
%%%%%%%%%%%%%%%%%%%%%%%%%%%%%%%%%%%%%%%%%%%%%%%%%%%%%%%%%%%%%%%%%%%%%%%%
%\begin{fmffile}{epgraphs}

\setcounter{footnote}{3}
%%%%%%%%%%%%%%%%%%%%%%%%%%%%%%%%%%%%%%%%%%%%%%%%%%%%%%%%%%%%%%%%%%%%%%%%
%%%%%%%%%%%%%%%%%%%%%%%%%%%%%%%%%%%%%%%%%%%%%%%%%%%%%%%%%%%%%%%%%%%%%%%%
\section{Introduction}
%%%%%%%%%%%%%%%%%%%%%%%%%%%%%%%%%%%%%%%%%%%%%%%%%%%%%%%%%%%%%%%%%%%%%%%%
%%%%%%%%%%%%%%%%%%%%%%%%%%%%%%%%%%%%%%%%%%%%%%%%%%%%%%%%%%%%%%%%%%%%%%%%

In electron-positron collider experiments at LEP and SLC, a large number of
Z bosons has been collected so that the detailed study of the decays of the
Z boson has been 
made possible \cite{lepwg}. An
interesting topic is the test of CP symmetry in
such Z decays.  There is already a number of theoretical 
(\cite{othertheo1}-\cite{over} and references therein) and experimental
\cite{alephtautau92}-\cite{opal}
studies of this subject. In the present paper we will study a
flavour-diagonal Z decay where
CP-violating effects within the
Standard Model (SM) are estimated to be very small \cite{zdecay}. Thus,
looking for CP violation in such Z decays
means looking for new physics beyond the SM. 

%In this paper we deal with the
%question: {\em Does CP violation beyond the SM exist in decays Z
%$\rightarrow$ 4 jets?}
%
For a model-independent systematic analysis of CP violation in Z decays
we use the effective Lagrangian approach as described in
\cite{zdecay,xsec}. Of particular interest are Z decays involving heavy
leptons or quarks. Thus, the process Z $\rightarrow b \bar{b} G$ which is
sensitive to effective CP-violating couplings in the $Z b \bar{b} G$ vertex
has been analysed theoretically in \cite{width,hab} and experimentally in
\cite{aleph}. No significant deviation from the SM has
been found. 

Here we present an analysis of the 4 jet decays of the Z boson
involving $b$ quarks. If CP-violating couplings are introduced in the $Z b
\bar{b}
G$ vertex, they will, because of gauge invariance of QCD, appear in the $Z
b \bar{b} G G$ vertex as well. But the $Z b \bar{b} G G$ vertex could in
principle contain new coupling parameters. The 4 jet analysis looks into
both, 4- and 5-point vertices.

In this paper we present the results of our calculations of the process Z
$\rightarrow$ 4 jets including CP-violating couplings, with at least two of
the jets originating from a $b$ or $\bar{b}$ quark. 
The following three subprocesses contribute to the 4 jet decay:
\begin{eqnarray}
  && e^+\,(p_+) \; e^-\,(p_-) 
\rightarrow Z\,(p) \rightarrow b\,(k_-)\;
\bar{b}\,(k_+)\; G\,(k_1)\; G\,(k_2) \, ,
\label{proc1}
\\[0.3cm]
&& e^+\,(p_+) \; e^-\,(p_-) 
\rightarrow Z\,(p) \rightarrow b\,(k_-)\;
\bar{b}\,(k_+)\; b\,(q_-)\; \bar{b}\,(q_+) \, ,
\label{proc2}
\\[0.3cm]
&& e^+\,(p_+) \; e^-\,(p_-) 
\rightarrow Z\,(p) \rightarrow b\,(k_-)\;
\bar{b}\,(k_+)\; q\,(q_-)\; \bar{q}\,(q_+)
\label{proc3}
\\
&& \hphantom{e^+\,(p_+) \; e^-\,(p_-) 
\rightarrow Z\,(p) \rightarrow b\,(k_-)\;
\bar{b}\,(k_+} (q= u, d, s, c) \, .
\nonumber
\end{eqnarray}

We will always assume unpolarized $e^+$, $e^-$ beams and show the results
for each process individually as well as the
results for the sum of them. In the experiments, of course, only the sum of
the three processes can be observed easily.

In chapter 2 we explain the theoretical framework of our
computations. Next, in chapter 3, we analyse the anomalous couplings for
partons in the final state. First, we discuss anomalous contributions to the
decay width. Then, we define different CP-odd tensor and
vector observables
and calculate their sensitivities to anomalous couplings. In order to find out
how ``good'' for the measurement of the new couplings our observables are, we
compare them to the optimal observables. In chapter 4 we study decay width,
tensor, vector and optimal observables in four different scenarios for an
experimental analysis. Finally, we compare our results with results of the
3 jet decay. Our conclusions can be found in chapter 5.

%%%%%%%%%%%%%%%%%%%%%%%%%%%%%%%%%%%%%%%%%%%%%%%%%%%%%%%%%%%%%%%%%%%%%%%%
%%%%%%%%%%%%%%%%%%%%%%%%%%%%%%%%%%%%%%%%%%%%%%%%%%%%%%%%%%%%%%%%%%%%%%%%
\section{Effective Lagrangian Approach}
%%%%%%%%%%%%%%%%%%%%%%%%%%%%%%%%%%%%%%%%%%%%%%%%%%%%%%%%%%%%%%%%%%%%%%%%
%%%%%%%%%%%%%%%%%%%%%%%%%%%%%%%%%%%%%%%%%%%%%%%%%%%%%%%%%%%%%%%%%%%%%%%%

For a model independent study of CP violation in 4 jet
decays of the Z boson we use the effective Lagrangian approach as explained
in \cite{zdecay}. We add to the SM Lagrangian ${\cal L}_{SM}$ a
CP-violating term ${\cal L}_{CP}$
containing all CP-odd local operators with a mass dimension $d \leq
6$ (\underline{after} electroweak symmetry breaking)
that can be constructed with SM fields. The effective CP-violating
Lagrangian relevant to our analysis is:
\begin{eqnarray}
  \label{lcp} \nonumber
  \lefteqn{ {\cal L}_{CP}(x) = } \\ \nonumber
  & - & \frac{i}{2} \tilde{d}_b \: \bar{b}(x)\: \sigma^{\mu\nu}\: \gamma_5\:
  b(x)\; [\partial_{\mu}\: Z_{\nu}(x) - \partial_{\nu}\: Z_{\mu}(x)] \\
  \nonumber 
  & - & \frac{i}{2} d_b' \: \bar{b}(x)\: T^a\: \sigma^{\mu\nu}\: \gamma_5\:
  b(x)\: G^a_{\mu\nu}(x) \\ 
  & + & [\; h_{Vb}\: \bar{b}(x)\: T^a\: \gamma^{\nu}\:
  b(x) +  h_{Ab}\: \bar{b}(x)\: T^a\: \gamma^{\nu}\: \gamma_5\: b(x)\; ]\;
  Z^{\mu}(x)\: 
  G^a_{\mu\nu}(x)\;\;,
\end{eqnarray}
where $b(x)$ denotes the b quark field, $Z^{\mu}(x)$ and $G^a_{\mu\nu}(x)$
represent the field of
the Z boson and the field strength tensor of the gluon, respectively, and
$T^a=\lambda^a/2$ are the 
generators of $SU(3)_C$ \cite{buch}. 
In (\ref{lcp}) $\tilde{d}_b$ is the weak dipole moment, $d_b'$ the
chromoelectric dipole moment, and  $h_{Vb}$, $h_{Ab}$ are CP-violating
vector and axial vector chirality conserving coupling constants.  
As effective coupling constants in ${\cal L}_{CP}$ the parameters
$\tilde{d}_b$, $d_b'$, $h_{Vb}$, $h_{Ab}$ are real. They are related to
form factors of vertices but should not be confused with the latter
(cf. e.~g. \cite{remarks}). 

Information on the spin of the final state partons in (\ref{proc1} --
\ref{proc3}) is hardly available experimentally. Thus, we consider as
observables only the parton's energies and momenta. Then, effects linear in the
dipole form factors $\tilde{d}_b$ and $d_b'$ are suppressed by
powers of $m_b/m_Z$. So angular
correlations of the jets in Z $\rightarrow$ 4 jets 
are only sensitive to the couplings $h_{Vb}$ and
$h_{Ab}$.

%%%%%%%%%%%%%%%%%%begin%%figure%%%%%%%%%%%%%%%%%%%%%%%%%%%%%%%%%%%%%%%
\begin{figure}[ht]
  \begin{center} \epsfig{file=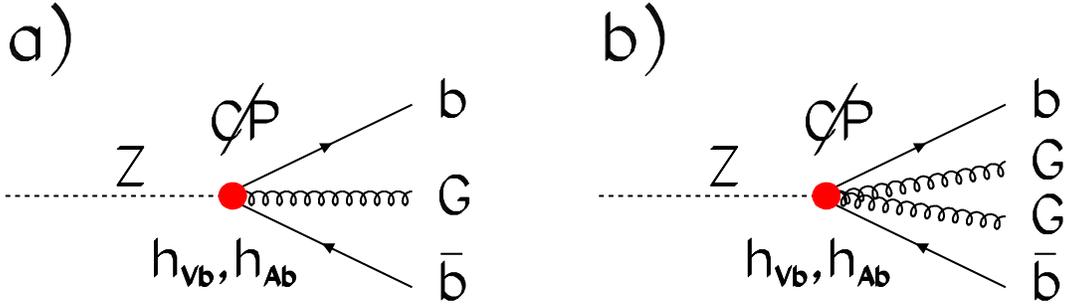,width=\hsize}
    \parbox{0.9\textwidth}
{\caption{\it{
The CP-violating vertices.
    }}\label{fig:vert}
}  \end{center}
\end{figure}
%%%%%%%%%%%%%%%%%%%%%%%%%%%%%%end%%figure%%%%%%%%%%%%%%%%%%%%%%%%%%%%%

The corresponding vertices following from ${\cal L}_{CP}$ are shown 
in figure~\ref{fig:vert}. Because the non-abelian field strength tensor has
a term 
quadratic in the gluon fields the $Z b \bar{b} G$- and $Z b \bar{b} G
G$-vertices are related.

We define dimensionless coupling constants $\hat{h}_{Vb,Ab}$ using the Z
mass as the scale parameter by
\begin{equation}
        h_{Vb,Ab} = \frac{e\: g_s}{\sin \vartheta_W\: \cos \vartheta_W \:m_Z^2}\; \hat{h}_{Vb,Ab}\;.
\end{equation}
For numerical calculations we set $m_Z=91.187\:$GeV, $\sin^2 \vartheta_W =
0.2236$ and the fine structure constant at the Z mass
to $\alpha =1/128.9$ \cite{pdg}. Our calculations are carried out in
leading order of the 
CP-violating couplings of
${\cal L}_{CP}$ and the SM couplings. A non-vanishing $b$ quark mass of
$4.5\:$GeV is included \footnote{We use here the pole mass value for the b
  quark. In our leading order calculation we could as well use the b mass
  at $m_Z$: $m_b(m_Z) = 2.83$ GeV \cite{mbatmz}. This would result only in
  minimal changes in our correlations.}
; masses of $u$, $d$, $s$, $c$ quarks are neglected.

%%%%%%%%%%%%%%%%%%%%%%%%%%%%%%%%%%%%%%%%%%%%%%%%%%%%%%%%%%%%%%%%%%%%%%%%
%%%%%%%%%%%%%%%%%%%%%%%%%%%%%%%%%%%%%%%%%%%%%%%%%%%%%%%%%%%%%%%%%%%%%%%%
\section{Study of CP-violating couplings for partons in the final state}
\label{sec:partons}
%%%%%%%%%%%%%%%%%%%%%%%%%%%%%%%%%%%%%%%%%%%%%%%%%%%%%%%%%%%%%%%%%%%%%%%%
%%%%%%%%%%%%%%%%%%%%%%%%%%%%%%%%%%%%%%%%%%%%%%%%%%%%%%%%%%%%%%%%%%%%%%%%

In this chapter we discuss an ideal experiment where one is able to
flavour-tag the partons and measure their momenta. We present a study
of our CP-violating couplings for each 
process (\ref{proc1}) -- (\ref{proc3}) separately and for the sum of them.
We have computed the differential and integrated decay rates using FORM
\cite{form} and M \cite{M} for the analytic and VEGAS \cite{vegas} 
for the numerical calculation. We write the squared matrix element for each
subprocess with final state $\PRO = {b \bar{b} G G},\; {b \bar{b} b
  \bar{b}},\; {b \bar{b} u 
  \bar{u}},\; {b \bar{b} d \bar{d}},\; {b \bar{b} s \bar{s}},\; {b \bar{b} c
  \bar{c}}$ in the form: 
\begin{eqnarray}
  \label{mezerl} \nonumber
  R(\phi)^{(\PRO)} &=& S_0^{(\PRO)}(\phi) \\ \nonumber
  &+& \hb S_1^{(\PRO)}(\phi) + \hbn S_2^{(\PRO)}(\phi) \\ 
  &+& (\hvb^2+\hab^2) S_3^{(\PRO)}(\phi) + (\hvb^2-\hab^2)
  S_4^{(\PRO)}(\phi) + \hvb\hab S_5^{(\PRO)}(\phi) \;.
\end{eqnarray}
Here $\phi$ stands collectively for the phase space variables,
$S_0$ denotes the SM part and 
\begin{equation}
\label{hat}
        \hat{h}_b = \hat{h}_{Ab} g_{Vb} - \hat{h}_{Vb} g_{Ab} \;,
\end{equation}
\begin{equation}
\label{tilde}
        \tilde{h}_b = \hat{h}_{Vb} g_{Vb} - \hat{h}_{Ab} g_{Ab} \;,
\end{equation}
\begin{equation}
  g_{Vb} = -\frac{1}{2} + \frac{2}{3} \sin^2 \vartheta_W \;, \;\; g_{Ab}=
  -\frac{1}{2} \;.
\end{equation}
In the following we drop the index \PRO\, if the given formula holds
for the subprocesses and for the sum of the subprocesses.

The results within the SM
have been compared analytically to calculations for vanishing b quark mass
\cite{reiter,reiternach} 
and to calculations for non-vanishing b quark mass \cite{arnd}. Our results
agree with these calculations.

The definition of a 4 jet sample requires the introduction of
resolution cuts. We use JADE cuts \cite{jade} requiring
\begin{equation}
        y_{ij} = \frac{2\, E_iE_j\,(1-\cos \vartheta_{ij})}{m_Z^2} >
        y_{cut} \;,
\label{jade}
\end{equation}
with $\vartheta_{ij}$ the angle between the momentum directions of any two
partons ($i \neq j$) and $E_i$, $E_j$ their energies in the Z rest
system.
The expectation value of an observable ${\cal{O}}(\phi)$ is then defined as
\begin{equation}
   <{\cal{O}}> = \frac{\int {\cal{O}} (\phi) \; R(\phi) \; d \phi}{\int
     R(\phi) \; d \phi} \, . 
\label{erwwobsallg}
\end{equation}

%%%%%%%%%%%%%%%%%%%%%%%%%%%%%%%%%%%%%%%%%%%%%%%%%%%%%%%%%%%%%%%%%%%%%%%%
\subsection{Anomalous contributions to the decay widths}
%%%%%%%%%%%%%%%%%%%%%%%%%%%%%%%%%%%%%%%%%%%%%%%%%%%%%%%%%%%%%%%%%%%%%%%%

\begin{figure}[H]
  \begin{center} \epsfig{file=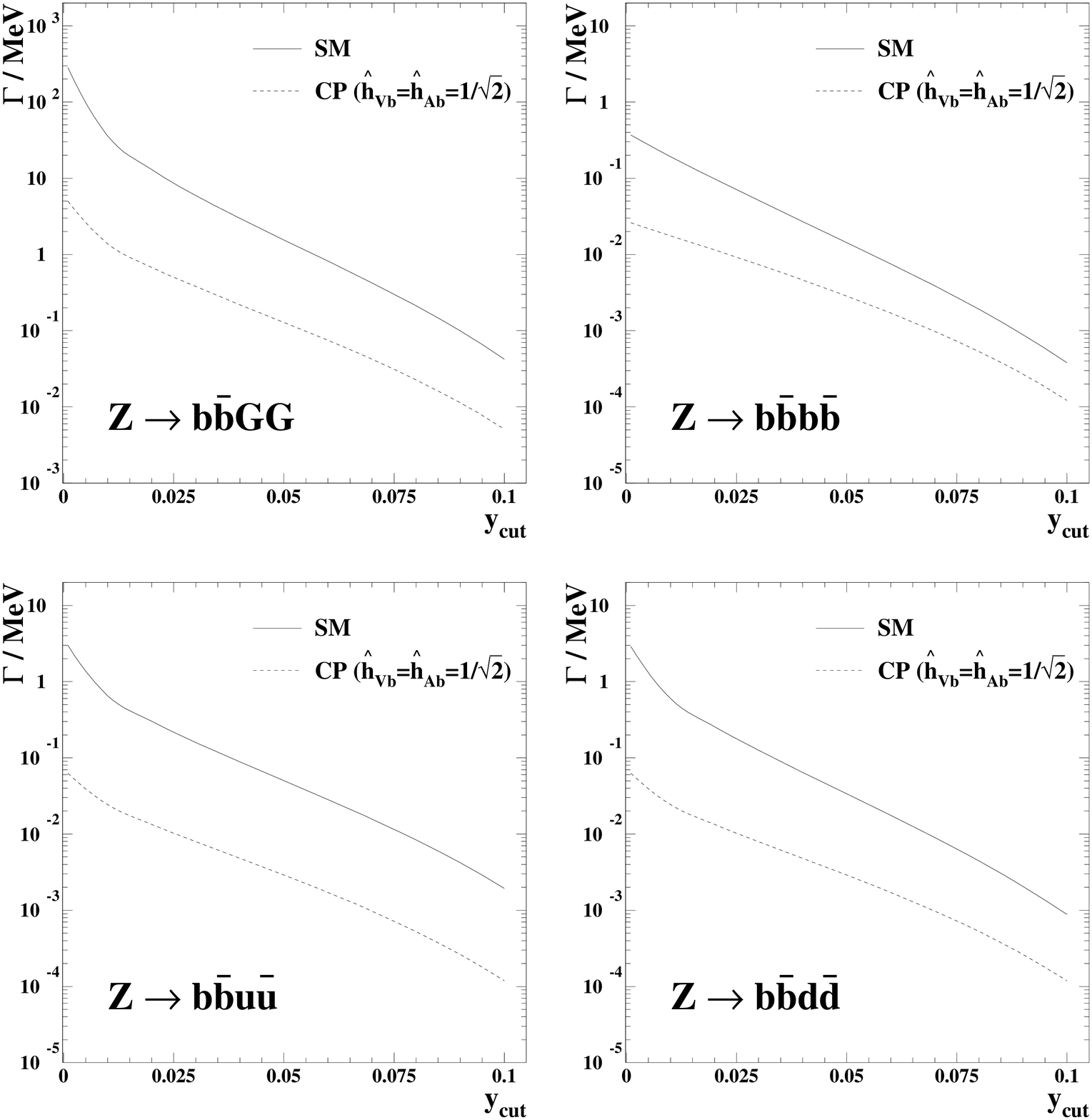,width=\hsize}
    \parbox{0.9\textwidth}{\caption{\it{
The decay width for different subprocesses as function of the jet
resolution parameter $y_{cut}$ (\ref{jade}). The results for {$Z
  \rightarrow b \bar{b} c \bar{c}$} ($b \bar{b} s \bar{s}$) are the same as
for \ZU\ ($b \bar{b} d \bar{d}$).
    }}\label{fig:widsub}
}  \end{center}
\end{figure}

The solid curves in figure~\ref{fig:widsub} show the results of our
calculations for the SM decay widths $\Gamma^{SM}$ as function of the jet
resolution parameter 
$y_{cut}$ for the different processes.
To check our calculations we computed $\Gamma^{SM}$ also with the program
COMPHEP \cite{comphep} and found --- within numerical errors --- complete
agreement. 

%%%%%%%%%%%%%%%%%%%begin%%figure%%%%%%%%%%%%%%%%%%%%%%%%%%%%%%%%%%%%%%%
%%%%%%%%%%%%%%%%%%%%%%%%%%%%%%end%%figure%%%%%%%%%%%%%%%%%%%%%%%%%%%%%

%%%%%%%%%%%%%%%%%%%begin%%figure%%%%%%%%%%%%%%%%%%%%%%%%%%%%%%%%%%%%%%%
\begin{figure}[H]
  \begin{center} \epsfig{file=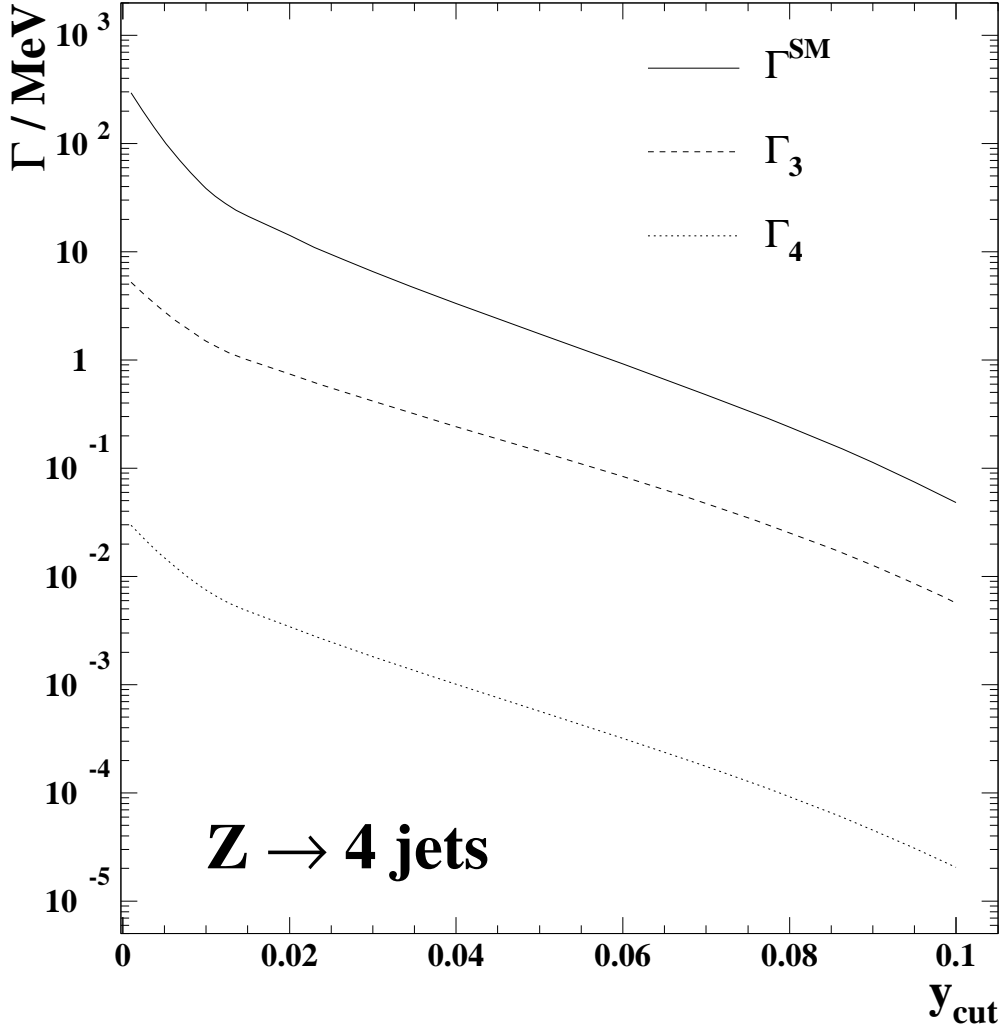,width=0.9\hsize}
    \parbox{0.9\textwidth}{\caption{\it{
The different contributions to the 4 jet decay width as function of the jet
resolution parameter for the sum of the processes (\ref{proc1} --
\ref{proc3}).
    }}\label{fig:widsmcp2}
}  \end{center}
\end{figure}
%%%%%%%%%%%%%%%%%%%%%%%%%%%%%%end%%figure%%%%%%%%%%%%%%%%%%%%%%%%%%%%%

As the decay width is a CP-even observable the contribution of the
CP-violating interaction to it adds incoherently to the SM one
\cite{width}:
\begin{equation}
        \Gamma = \Gamma^{SM} + \Delta\Gamma^{CP} \;,
\end{equation}
with $\Delta\Gamma^{CP}$ being quadratic in the
new couplings. In figure~\ref{fig:widsub}
the dashed curves represent $\Delta\Gamma^{CP}$ as function of $y_{cut}$
assuming $\hat{h}_{Vb}=\hat{h}_{Ab}=1/\sqrt{2}$. 

As we can see, the dominant decay is (\ref{proc1}). In
comparison to this process, the
processes (\ref{proc3}) give only contributions at the per cent level,
process (\ref{proc2}) at the per mille level to the decay width.
From (\ref{mezerl}) we find:
\begin{equation}
\label{gamzerl}
\Delta\Gamma^{CP} = (\hvb^2+\hab^2) \Gamma_3 + (\hvb^2-\hab^2)
\Gamma_4 \;.
\end{equation}
Because $S_5$
in (\ref{mezerl}) turns out to be odd under the exchange of quark and
anti-quark momenta, its integral vanishes.
In Figure~\ref{fig:widsmcp2}, we compare 
for the sum of the processes (\ref{proc1} -- \ref{proc3}) $\Gamma^{SM}$,
$\Gamma_3$ and $\Gamma_4$.
For $\hat{h}_{Vb}$, $\hat{h}_{Ab}$ of order one
$\Delta\Gamma^{CP}$ is only a correction of a few per cent to
$\Gamma^{SM}$ even if all processes (\ref{proc1} -- \ref{proc3}) are added
up. 
Thus, considering the theoretical uncertainties in the SM 4 jet decay rate,
a determination of the new couplings
by measuring the decay width alone does not look promising.
 
%%%%%%%%%%%%%%%%%%%%%%%%%%%%%%%%%%%%%%%%%%%%%%%%%%%%%%%%%%%%%%%%%%%%%%%%
\subsection{CP-odd observables}
%%%%%%%%%%%%%%%%%%%%%%%%%%%%%%%%%%%%%%%%%%%%%%%%%%%%%%%%%%%%%%%%%%%%%%%%

%%%%%%%%%%%%%%%%%%%%%%%%%%%%%%%%%%%%%%%%%%%%%%%%%%%%%%%%%%%%%%%%%%%%%%%%
\subsubsection{Tensor and vector observables}
\label{sec:parttenvec}
%%%%%%%%%%%%%%%%%%%%%%%%%%%%%%%%%%%%%%%%%%%%%%%%%%%%%%%%%%%%%%%%%%%%%%%%

We now turn to a study of our CP-violating couplings using
CP-odd observables constructed from the
momentum directions of the $b$ and $\bar{b}$ quarks,
$\hat{\bf k}_b={\bf k}_b/|{\bf k}_b|$  and $\hat{\bf k}_{\bar{b}}={\bf k}_{\bar{b}}/|{\bf k}_{\bar{b}}|$
(cf. \cite{zdecay,xsec,bernnach2,hab}):
\begin{equation}
        T_{ij}^{(n)} = (\hat{\bf k}_{\bar{b}} - \hat{\bf k}_b)_i \; (\hat{\bf k}_{\bar{b}} \times
\hat{\bf k}_b)_j \; |\hat{\bf k}_{\bar{b}} \times \hat{\bf k}_b|^{n-2} + (i \leftrightarrow j) \; ,
\label{ten}
\end{equation}
\begin{equation}
        V_i^{(n)} = (\hat{\bf k}_{\bar{b}} \times \hat{\bf k}_b)_i \; |\hat{\bf k}_{\bar{b}} \times
\hat{\bf k}_b|^{n-2} \; ,
\label{vec}
\end{equation}
with $i$, $j$ the Cartesian vector indices in the Z rest system and
$n=1,2,3$.
 
The observables $T_{ij}^{(n)}$ transform as tensors, $V_i^{(n)}$ as
vectors. For unpolarized $e^+e^-$ beams and our rotationally invariant cuts
(\ref{jade}) 
their expectation values are then proportional to the Z
tensor polarization $S_{ij}$ and vector polarization $s_i$,
respectively. Defining the positive $z$-axis in the $e^+$ beam
direction, we have
\begin{equation}
  {\bf s} = \left( \begin{array}{c}
  0 \\ 0 \\ s_3
\end{array} \right) \;,
\end{equation}
\begin{equation}
  (s_{ij}) = \frac{1}{6} \left( \begin{array}{ccc}
  -1 & 0 & 0 \\
  0 &-1 & 0 \\
  0 & 0 & 2    
\end{array} \right) \;,
\end{equation}
where 
\begin{equation}
        s_3 = \frac{2 \, g_{Ve} g_{Ae}}{g_{Ve}^2 + g_{Ae}^2} = 0.209\;,
\end{equation}
with $g_{Ve}=-1/2+2 \sin^2 \vartheta_W$ and $g_{Ae}=-1/2$ the weak vector
and axial vector $Zee$
couplings. This shows that the components $T_{33}^{(n)}$ and $V_3^{(n)}$
are the most sensitive ones.

Note that the tensor observables do {\em not} change their sign upon charge
misidentification ($\hat{\bf k}_{\bar{b}} \leftrightarrow \hat{\bf k}_b$) whereas the
vector observables do. Thus, it is only for the measurement of the latter
that charge identification is indispensable, which makes the vector
observables less valuable for the experimental analysis.

We have computed the expectation values of the observables (\ref{ten}),
(\ref{vec}) for different JADE cuts (\ref{jade}), as function of
$\hat{h}_b$ (\ref{hat}) and $\tilde{h}_b$ (\ref{tilde}).
The expectation value of a CP-odd observable ${\cal{O}}$
has the following general form:
\begin{equation}
        <\!{\cal{O}}\!> = (c_1 \hat{h}_b + c_2 \tilde{h}_b) \,
        \frac{\Gamma^{SM}_{4\;jets}}{\Gamma_{4\;jets}} \;,
\label{cpoddobs}
\end{equation}
where $\Gamma^{SM}_{4\;jets}$ and $\Gamma_{4\;jets}$ denote the
corresponding \ZJ\
decay widths in the SM and in the theory with SM plus CP-violating couplings,
respectively. 
In an experimental analysis $\Gamma^{SM}_{4\;jets}$ should be
taken from the theoretical calculation, $\Gamma_{4\;jets}$ and
$<\!{\cal{O}}\!>$ from the experimental measurement. The quantity
$<\!{\cal{O}}\!> \! \cdot \,\Gamma_{4\;jets}$ is then an observable strictly linear in
the anomalous couplings.

From the measurement of a single observable (\ref{cpoddobs}) we can get a
simple estimate of its sensitivity to \hb\ by assuming $\hbn=0$. The
error on a measurement of \hb\ is then to leading order in the anomalous
couplings:
\begin{equation}
        \delta\hat{h}_b =
        \frac{\sqrt{<\!{\cal{O}}^2\!>_{SM}}}{|c_1| \sqrt{N}} \;,
\label{dhb}
\end{equation}
where $N$ is the number of events within cuts. Similarly, assuming
$\hb=0$ we get the error on \hbn\ as 
\begin{equation}
        \delta\tilde{h}_b =
        \frac{\sqrt{<\!{\cal{O}}^2\!>_{SM}}}{|c_2| \sqrt{N}} \;.
\label{dhbn}
\end{equation}
A measure for the sensitivity of \OBS\ to \hb\ (\hbn) is then $1/\dhb$
($1/\dhbn$).
However, since we want to estimate 2 anomalous couplings \hb, \hbn\ we
should consider 2 linearly independent observables $\OBS_{1,2}$ such that:
\begin{eqnarray} \nonumber
  <\! \OBS_1 \!> = ( \hb c_{11} + \hbn c_{12} ) \,
        \frac{\Gamma^{SM}_{4\;jets}}{\Gamma_{4\;jets}} \;, \\
  <\! \OBS_2 \!> = ( \hb c_{21} + \hbn c_{22} ) \,
        \frac{\Gamma^{SM}_{4\;jets}}{\Gamma_{4\;jets}} \;.
\label{evobs12}
\end{eqnarray}
The sensitivity of these observables to the anomalous
couplings is estimated in the standard way.
Neglecting terms quadratic in
the anomalous couplings the combined measurement of $<\! \OBS_1 \!>$ and $<\!
\OBS_2 \!>$ with a data sample of $N$ events (within the considered cuts)
leads to an error ellipse
\begin{equation}
\label{1sdellipse}
  (\delta\hb)^2 \, V(h)^{-1}_{11} + 2 \, \delta\hb \delta\hbn \,
  V(h)^{-1}_{12} + (\delta\hbn)^2 \, V(h)^{-1}_{22} = 1 \;.
\end{equation}
Here ${\sf V}(h)$ denotes the covariance matrix of the estimated couplings.
We have in matrix notation:
\begin{equation}
  {\sf V}(h) = \frac{1}{N} {\sf c}^{-1}{\sf V}(\OBS) ({\sf c}^{-1})^T \;,
\end{equation}
\begin{equation}
\label{coeffobs}
  {\sf c} = \left( \begin{array}{cc}
  c_{11} & c_{12} \\
  c_{21} & c_{22} 
\end{array} \right) \;,
\end{equation}
where
\begin{equation}
\label{covobs}
  V(\OBS)_{ij} =  \frac{1}{\int S_0(\phi) d\phi}\int \OBS_i(\phi)
  \OBS_j(\phi) S_0(\phi) d\phi
\end{equation}
are the elements of the covariance matrix of the observables $\OBS_i$,
calculated in the SM. A measurement of \hb , \hbn\ has to produce a mean
value point outside the ellipse (\ref{1sdellipse}) to be able to claim a
non-zero effect at the 1 s.~d. level. 

%%%%%%%%%%%%%%%%%%%%%%%%%%%%%%%%%%%%%%%%%%%%%%%%%%%%%%%%%%%%%%%%%%%%%%%%
\subsubsection{Optimal observables}
%%%%%%%%%%%%%%%%%%%%%%%%%%%%%%%%%%%%%%%%%%%%%%%%%%%%%%%%%%%%%%%%%%%%%%%%
\label{sec:partopt}

In addition to the tensor and vector observables (\ref{ten}, \ref{vec})
we study 
{\em optimal observables}, which have the largest possible statistical
signal-to-noise ratio \cite{opt,opt1,opt2}. Neglecting higher orders 
in the anomalous couplings the optimal
observables for measuring \hb\ and \hbn\ are obtained from the differential
cross section (\ref{mezerl}) as
\begin{equation}
  O_i(\phi) = \frac{S_i(\phi)}{S_0(\phi)} \;, \;\;\;\;(i=1,2) \;.
\label{opti}
\end{equation}
The expectation values for the optimal observables are as in
(\ref{evobs12}) with the coefficient matrix elements
\begin{equation}
\label{cijopt}
  c_{ij} = \frac{1}{\int S_0 d\phi}
  \int \frac{S_i(\phi)S_j(\phi)}{S_0(\phi)}d\phi \;.
\end{equation}
For optimal observables we have
\begin{equation}
  c_{ij} = V(\OBS)_{ij} \;,
\end{equation}
\begin{equation}
  V(h)^{-1}_{ij} = N c_{ij} \;.
\end{equation}

%%%%%%%%%%%%%%%%%%%%%%%%%%%%%%%%%%%%%%%%%%%%%%%%%%%%%%%%%%%%%%%%%%%%%%%%
\subsubsection{Numerical results}
%%%%%%%%%%%%%%%%%%%%%%%%%%%%%%%%%%%%%%%%%%%%%%%%%%%%%%%%%%%%%%%%%%%%%%%%
\label{sec:partres}

We have calculated the sensitivities to
$\hat{h}_b$ and $\tilde{h}_b$ 
for different tensor, vector and the optimal
observables varying the jet resolution parameter \ycut\ .
We assume a total number of $10^4$ 4 jet events from (\ref{proc1}
-- \ref{proc3}) for $y_{cut}=0.01$:
\begin{equation}
  \label{nevents}
  N(y_{cut}=0.01)=10000 \;.
\end{equation}
The number of events for other values
of $y_{cut}$ and for the various subprocesses is then calculated within
the SM. The total number of Z decays corresponding to (\ref{nevents}) is
$N_{tot} \cong 6.4 \cdot 10^5$.

\begin{figure}[H]
  \begin{center} \epsfig{file=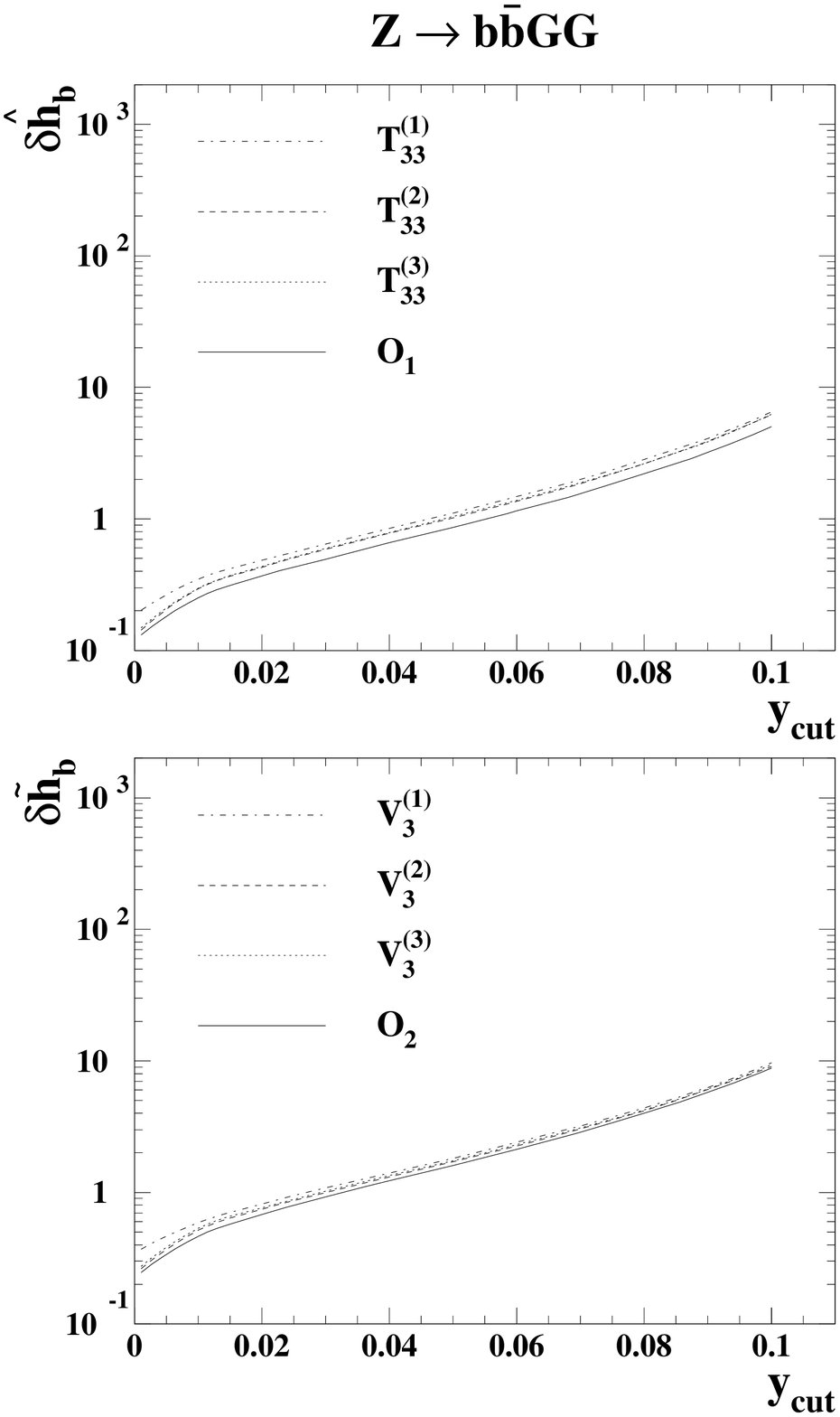,width=0.75\hsize}
    \parbox{0.9\textwidth}
{\caption{\it{
The inverse sensitivities of tensor, vector and optimal observables to
$\hat{h}_b$ and $\tilde{h}_b$ (\ref{hat},\ref{tilde}) obtainable in \ZG,
as function of the jet resolution parameter $y_{cut}$ (\ref{jade}) assuming
(\ref{nevents}) for the number of events.
    }}\label{fig:obsgg}
}  \end{center}
\end{figure}

For the process $Z \rightarrow b \bar{b} G G$, we found that, in very good
approximation, the tensor observables are only sensitive to
$\hat{h}_b$ and the vector observables only to 
$\tilde{h}_b$. The sensitivities to these
CP-odd couplings as calculated from (\ref{dhb}) and (\ref{dhbn}),
respectively, are shown in figure~\ref{fig:obsgg}.
The sensitivity decreases with increasing $y_{cut}$ for all
observables due to the decrease in number of events available. The
differences due to the different weight factors for tensor and
vector observables $T_{33}^{(n)}$, $V_3^{(n)}$ ($n=1,2,3$) are only small
and all observables considered have nearly optimal sensitivities. 

\begin{figure}[H]
  \begin{center} \epsfig{file=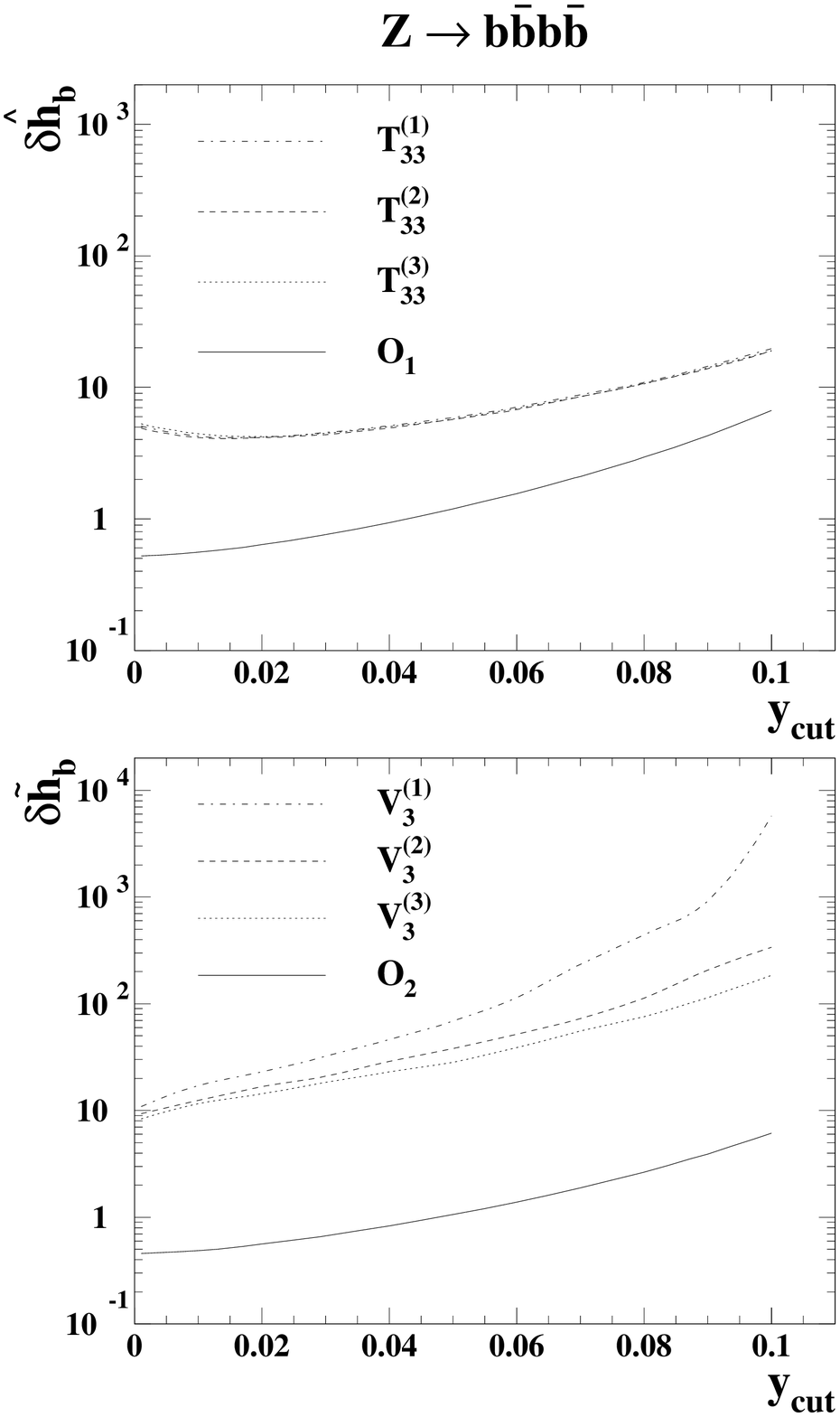,width=0.75\hsize}
    \parbox{0.9\textwidth}
{\caption{\it{
The inverse sensitivities of tensor, vector and optimal observables to
$\hat{h}_b$ and $\tilde{h}_b$ (\ref{hat},\ref{tilde}) obtainable in \ZB,
as function of the jet resolution parameter $y_{cut}$ (\ref{jade}) assuming
(\ref{nevents}) for the number of events.
    }}\label{fig:obsbb}
}  \end{center}
\end{figure}

\begin{figure}[H]
  \begin{center} \epsfig{file=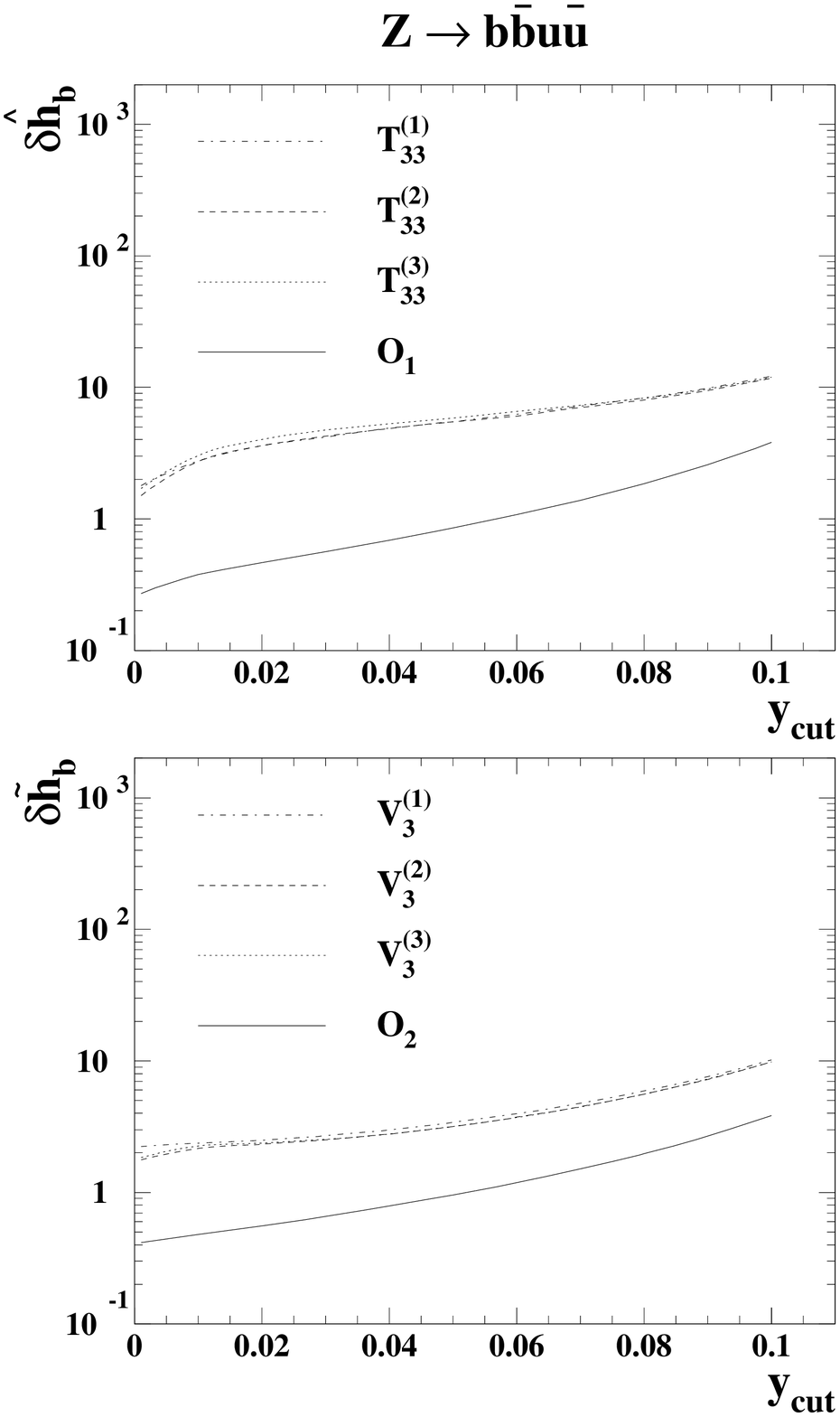,width=0.75\hsize}
    \parbox{0.9\textwidth}
{\caption{\it{
The inverse sensitivities of tensor, vector and optimal observables to
$\hat{h}_b$ and $\tilde{h}_b$ (\ref{hat},\ref{tilde}) obtainable in \ZU,
as function of the jet resolution parameter $y_{cut}$ (\ref{jade}) assuming
(\ref{nevents}) for the number of events. The results for $Z
  \rightarrow b \bar{b} c \bar{c}$ are identical.
    }}\label{fig:obsuu}
}  \end{center}
\end{figure}

\begin{figure}[H]
  \begin{center} \epsfig{file=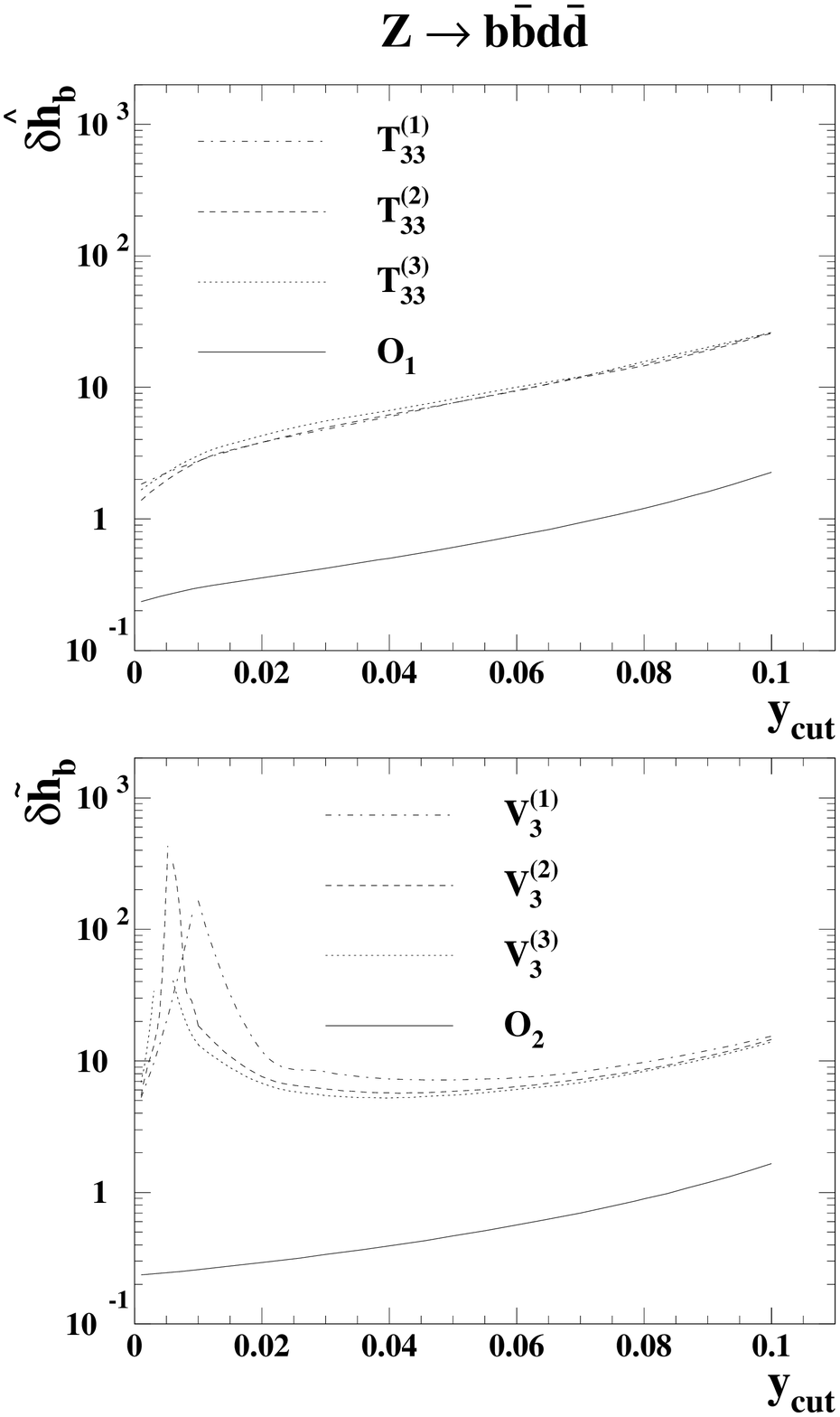,width=0.75\hsize}
    \parbox{0.9\textwidth}
{\caption{\it{
The inverse sensitivities of tensor, vector and optimal observables to
$\hat{h}_b$ and $\tilde{h}_b$ (\ref{hat},\ref{tilde}) obtainable in \ZD,
as function of the jet resolution parameter $y_{cut}$ (\ref{jade}) assuming
(\ref{nevents}) for the number of events. The results for $Z
  \rightarrow b \bar{b} s \bar{s}$ are identical.
    }}\label{fig:obsdd}
}  \end{center}
\end{figure}

For the processes $Z \rightarrow b \bar{b} b \bar{b}$, $Z \rightarrow b
\bar{b} u \bar{u}$ and $Z \rightarrow b \bar{b} d \bar{d}$ we present plots
analogous to those for $Z \rightarrow b \bar{b} G G$
in figures~\ref{fig:obsbb}
-- \ref{fig:obsdd}. However, here the correlation of the sensitivity of
\hb\ (\hbn) with tensor (vector) observables no longer holds.
In $Z \rightarrow b \bar{b} d \bar{d}$, for instance, the vector
observables are more sensitive to $\hat{h}_b$ than to $\tilde{h}_b$. 
In the bottom of figure \ref{fig:obsdd} we get a singularity for \dhbn\
because at $y_{cut} \approx 0.06$ the expectation values of our vector
observables become zero. Left
from the singularity all expectation values are negative, right from the
singularity they are positive. For the sum of the processes the singularity
in \hbn\ vanishes, because we have to add up both the variances of all
subprocesses in the denominator and the expectation values of all
subprocesses in the numerator of (\ref{dhbn}). 

Two results are striking:
\begin{enumerate}
\item The sensitivities obtainable with optimal observables from the
  subprocesses (\ref{proc2}, 
  \ref{proc3}) are as good as or even better than those from (\ref{proc1})
  even if 
  the number of events from (\ref{proc2}, \ref{proc3}) represents only a
  small fraction of those from (\ref{proc1}) (cf. figure~\ref{fig:widsub}).
\item For the processes (\ref{proc2}, \ref{proc3}) the tensor and vector
  observables do not reach optimal sensitivities.
\end{enumerate}

One may understand these points in the following way: In the Feynman graph
giving the CP-violating amplitude for process 
(\ref{proc3}), the gluon which comes out of the CP-violating vertex
splits into a $q$ $\bar{q}$ pair. This means that the CP-violating
vertex can be analysed not only by using the angular correlations of $b$
and $\bar{b}$ quark, but also, by means of the 
momentum directions of $q$ and $\bar{q}$ quarks.
If the tensor and vector observables (\ref{ten}, \ref{vec})
are used for the measurement all information on the
CP-violating vertex delivered from the angular correlations of 
$q$ and $\bar{q}$ quarks is lost but, on the other hand, it is retained by
the optimal observables.
For process (\ref{proc2}) this argumentation is similar.

\begin{figure}[H]
  \begin{center} \epsfig{file=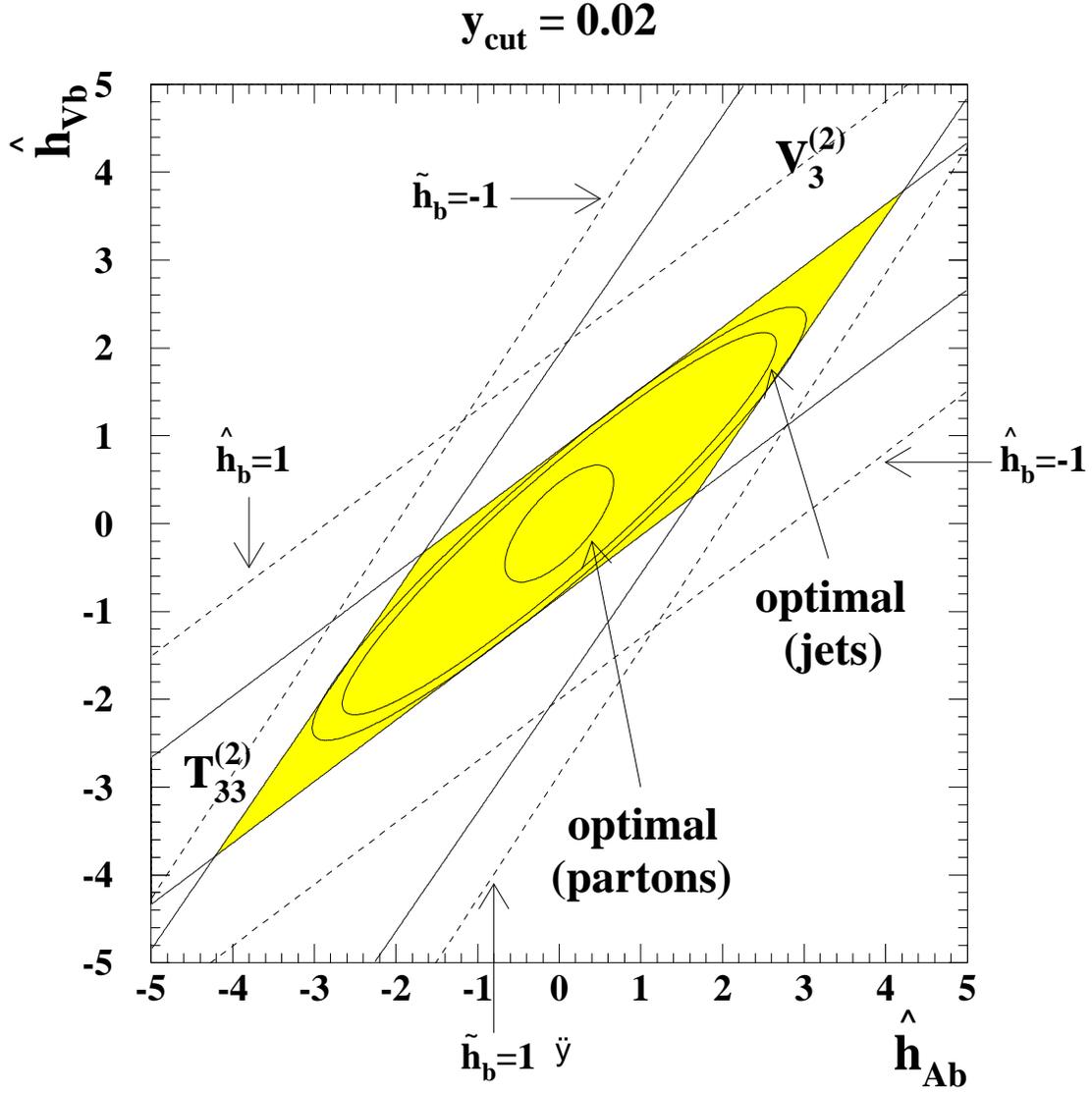,width=0.9\hsize}
    \parbox{0.9\textwidth}
{\caption{\it{
Contour plot for $y_{cut}=0.02$ of the 1 s.~d. errors on $\hat{h}_{Vb}$,
$\hat{h}_{Ab}$ as they can be obtained ideally from the measurement of
different observables for \ZJ. The innermost ellipsis is the result for the
optimal observables $O_1$ and $O_2$ at parton level. The outermost ellipsis
is the result for the combined measurement of $T_{33}^{(2)}$, $V_3^{(2)}$
at parton level. The solid bands correspond to a measurement of
$T_{33}^{(2)}$ or $V_3^{(2)}$ alone, respectively. The overlap region of
the tensor and 
vector errors is marked grey. The dashed lines correspond to $\hb = \pm 1$
and $\hbn = \pm 1$ as indicated.
The ellipsis in the middle is
the combined result for the optimal 
observables $O_1$ and $O_2$ for analysis 1 (cf. chapter~\ref{sec:jets}).
    }}\label{fig:sens}
}  \end{center}
\end{figure}

In figure~\ref{fig:sens}, results for the sum of the subprocesses
(\ref{proc1} -- \ref{proc3}) are shown for $y_{cut}=0.02$ for a combined
measurement of $T_{33}^{(2)}$ (\ref{ten}) and $V_3^{(2)}$ (\ref{vec}) and
for the optimal observables. The comparison of the solid bands (measurement of
$T_{33}^{(2)}$ or $V_3^{(2)}$ alone, respectively) with the
dashed lines corresponding to $\hb= \pm 1$ and $\hbn = \pm 1$ shows that
$T_{33}^{(2)}$ is mostly sensitive to \hb\ and that $V_3^{(2)}$
is mostly sensitive to \hbn.
The comparison of the inner- and outermost ellipses shows that tensor and 
vector observables do not reach the optimal sensitivity for the sum of the
processes (\ref{proc1} -- \ref{proc3}). This is remarkable since for the
dominant 
process (\ref{proc1}) they do. Thus, as already discussed above,
(\ref{proc2}) and (\ref{proc3}) which contribute little in the decay rate
have a much larger influence in CP-odd observables. In
tables~\ref{tab:coeffobs},~\ref{tab:covobs},~\ref{tab:cijopt} of
appendix~\ref{sec:numvalues}, we list the elements of the coefficient
matrix (\ref{coeffobs}) and the covariance matrix (\ref{covobs}) of the
observables $\OBS_1=T_{33}^{(2)}$ (\ref{ten}), $\OBS_2=V_{3}^{(2)}$
(\ref{vec}) and the coefficient matrix elements (\ref{cijopt}) for the
optimal observables (\ref{opti}) for different values of the jet resolution
parameter $y_{cut}$ (\ref{jade}).

%%%%%%%%%%%%%%%%%%begin%%figure%%%%%%%%%%%%%%%%%%%%%%%%%%%%%%%%%%%%%%%
%%%%%%%%%%%%%%%%%%%%%%%%%%%%%%end%%figure%%%%%%%%%%%%%%%%%%%%%%%%%%%%%

%%%%%%%%%%%%%%%%%%begin%%figure%%%%%%%%%%%%%%%%%%%%%%%%%%%%%%%%%%%%%%%
%%%%%%%%%%%%%%%%%%%%%%%%%%%%%%end%%figure%%%%%%%%%%%%%%%%%%%%%%%%%%%%%

%%%%%%%%%%%%%%%%%%begin%%figure%%%%%%%%%%%%%%%%%%%%%%%%%%%%%%%%%%%%%%%
%%%%%%%%%%%%%%%%%%%%%%%%%%%%%%end%%figure%%%%%%%%%%%%%%%%%%%%%%%%%%%%%

%%%%%%%%%%%%%%%%%%begin%%figure%%%%%%%%%%%%%%%%%%%%%%%%%%%%%%%%%%%%%%%
%%%%%%%%%%%%%%%%%%%%%%%%%%%%%%end%%figure%%%%%%%%%%%%%%%%%%%%%%%%%%%%%

%%%%%%%%%%%%%%%%%%begin%%figure%%%%%%%%%%%%%%%%%%%%%%%%%%%%%%%%%%%%%%%
%%%%%%%%%%%%%%%%%%%%%%%%%%%%%%end%%figure%%%%%%%%%%%%%%%%%%%%%%%%%%%%%

%%%%%%%%%%%%%%%%%%%%%%%%%%%%%%%%%%%%%%%%%%%%%%%%%%%%%%%%%%%%%%%%%%%%%%%%
%%%%%%%%%%%%%%%%%%%%%%%%%%%%%%%%%%%%%%%%%%%%%%%%%%%%%%%%%%%%%%%%%%%%%%%%
\section{CP-violating observables for jets}
\label{sec:jets}
%%%%%%%%%%%%%%%%%%%%%%%%%%%%%%%%%%%%%%%%%%%%%%%%%%%%%%%%%%%%%%%%%%%%%%%%
%%%%%%%%%%%%%%%%%%%%%%%%%%%%%%%%%%%%%%%%%%%%%%%%%%%%%%%%%%%%%%%%%%%%%%%%

In an experimental analysis one can only measure
jets as the ``footprints'' of the underlying partons, but not the partons
themselves. 
So if we want to compare our calculations directly to experimental data, we
must define observables for jets. In LEP experiments it is possible to tag
a jet referring to a quark with b flavour \cite{aleph}. In principle one
can even distinguish between $b$ and $\bb$ by measuring the jet
charge, but this is difficult in practice.
We propose four different types of analyses with the 4 jet data sample:
\begin{itemize}
\item {\bf Analysis 1}: One jet comes from $b$ fragmentation, another from
  $\bar{b}$ fragmentation (double b tag); the other two jets (jets 3 and 4)
  are ordered according to the magnitude of their momenta. 
\end{itemize}
For the
next three analyses, we
propose to make an ordering of all four jets according to the magnitude of
their momenta:
\begin{equation}
  |{\bf q}_1| \geq |{\bf q}_2| \geq |{\bf q}_3| \geq |{\bf q}_4| \;.
\label{momorder}
\end{equation}
In the following we call jet 1 the jet with the highest magnitude of
momentum, jet 2 the jet with the second highest magnitude of momentum and
so on. 
\begin{itemize}
\item {\bf Analysis 2}: Jet 1 comes from $b$ or $\bar{b}$ fragmentation.
\item {\bf Analysis 3}: Jet 2 comes from $b$ or $\bar{b}$ fragmentation.
\item {\bf Analysis 4}: No requirement to the jet flavour (flavour blind
  case).
\end{itemize}
In appendix \ref{sec:eventclass} we list the different classes of
events for each of the subprocesses (\ref{proc1} -- \ref{proc3}) as they
contribute to these analyses.

In analyses 2 -- 4 we do not distinguish between $b$ and \bb\ jets. It
turns out that this in essence eliminates the dependence of the distributions on the
CP-odd parameter \hbn. Thus here 
we can only measure \hb\ and we set $\hbn=0$ for these analyses.

%%%%%%%%%%%%%%%%%%%%%%%%%%%%%%%%%%%%%%%%%%%%%%%%%%%%%%%%%%%%%%%%%%%%%%%%
\subsection{Anomalous contributions to the decay width}
%%%%%%%%%%%%%%%%%%%%%%%%%%%%%%%%%%%%%%%%%%%%%%%%%%%%%%%%%%%%%%%%%%%%%%%%

\begin{figure}[H]
  \begin{center} \epsfig{file=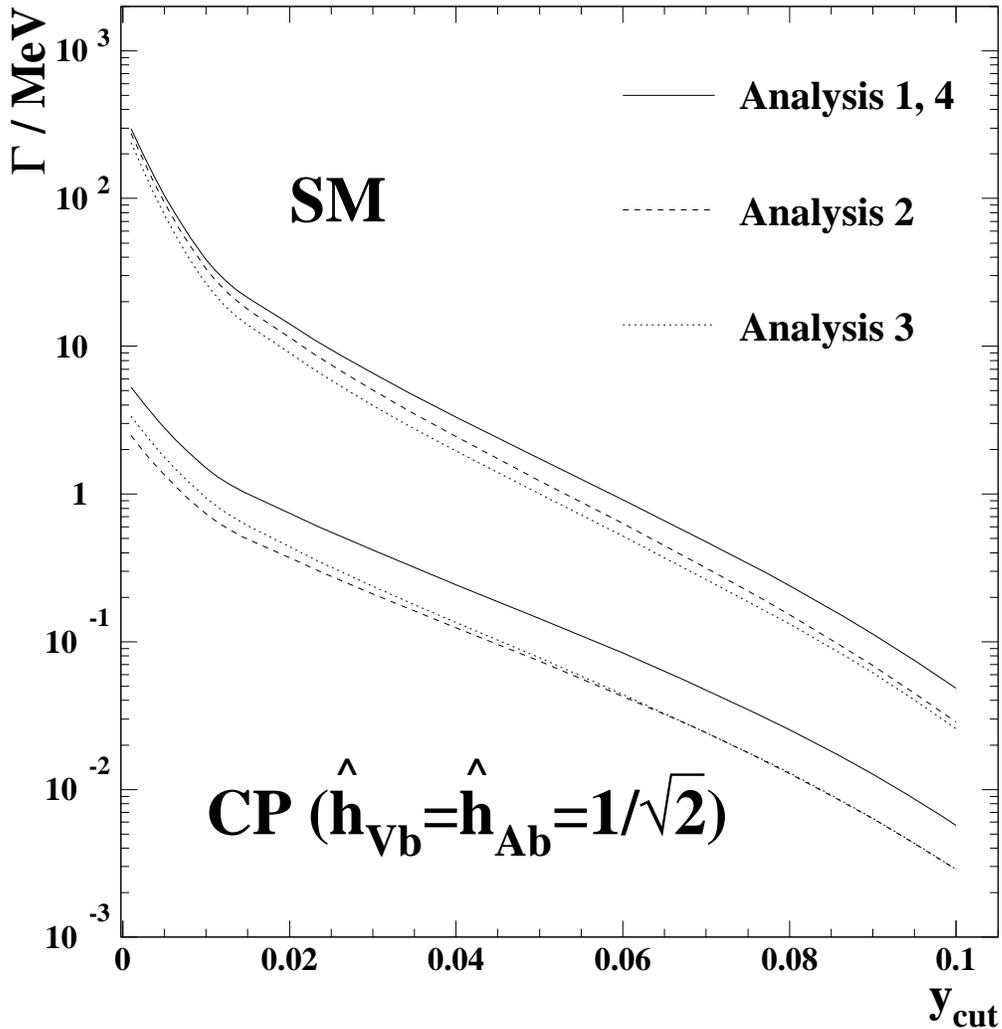,width=0.9\hsize}
    \parbox{0.9\textwidth}{\caption{\it{
The 4 jet decay width as function of the jet resolution parameter for
different analyses. The three upper curves are the SM contributions, the
others are the contributions from the CP-violating interaction if we set
$\hvb=\hab=1/\sqrt{2}$.
    }}\label{fig:wid}
}  \end{center}
\end{figure}

We computed the total decay width for the 4 jet decays of the Z boson with
at least two jets coming from $b$ or $\bar{b}$ fragmentation for the
different analyses.
Because a momentum ordering of jets can't influence a decay rate, the
analyses 1, 4 give results identical to those for the
partons in the final state. In
analyses 2, 3 some events are rejected as can be seen from the tables
\ref{tab:eventclassg}, \ref{tab:eventclassb},
\ref{tab:eventclassq} of appendix~\ref{sec:eventclass}. The decay width must decrease in comparison to the
other two analyses. Figure~\ref{fig:wid}
shows this effect both for the SM contribution and for
the contribution of the CP violating interaction   
to the decay width assuming $\hvb=\hab=1/\sqrt{2}$.

%%%%%%%%%%%%%%%%%%%begin%%figure%%%%%%%%%%%%%%%%%%%%%%%%%%%%%%%%%%%%%%%
%%%%%%%%%%%%%%%%%%%%%%%%%%%%%%end%%figure%%%%%%%%%%%%%%%%%%%%%%%%%%%%%

%%%%%%%%%%%%%%%%%%%%%%%%%%%%%%%%%%%%%%%%%%%%%%%%%%%%%%%%%%%%%%%%%%%%%%%%
\subsection{CP-odd observables}
%%%%%%%%%%%%%%%%%%%%%%%%%%%%%%%%%%%%%%%%%%%%%%%%%%%%%%%%%%%%%%%%%%%%%%%%

%%%%%%%%%%%%%%%%%%%%%%%%%%%%%%%%%%%%%%%%%%%%%%%%%%%%%%%%%%%%%%%%%%%%%%%%
\subsubsection{Tensor and vector observables}
%%%%%%%%%%%%%%%%%%%%%%%%%%%%%%%%%%%%%%%%%%%%%%%%%%%%%%%%%%%%%%%%%%%%%%%%

We found in chapter \ref{sec:partons} that the observables
$T_{33}^{(2)}$ (\ref{ten}) and $V_3^{(2)}$ (\ref{vec}) were the most
sensitive ones. The same was found for 
the 3 jet decays (cf. \cite{hab}).
Thus, from now on we concentrate on this type of observables.

\paragraph{Analysis 1:} The tensor and vector observables in this analysis
are the same as for partons:
$T_{33}^{(2)}$ (\ref{ten}) and $V_3^{(2)}$ (\ref{vec}). All results are
identical to 
the parton case summed over the subprocesses (\ref{proc1} --
\ref{proc3}) of chapter \ref{sec:partons}. 
Thus the sensitivity of a measurement of $T_{33}^{(2)}$ and $V_3^{(2)}$ 
to $\hat{h}_{Vb,Ab}$ for $y_{cut}=0.02$ is obtained from
figure~\ref{fig:sens}.
For the measurement of
the tensor observable, which is C-even, we do not need to distinguish between
\bb\ and $b$ quark. A sufficient selection criterion is then that we demand
two jets coming from $b$ \underline{or} \bb\ fragmentation. For the
measurement of the vector observable, which is C-odd, we need to distinguish
between jets coming from $b$ or \bb\ fragmentation. This can be done
experimentally by measuring the jet charge.

\paragraph{Analysis 2, 3, 4:} As tensor observable we chose now
\begin{equation}
        T{'}_{ij}^{(2)} = (\hat{\bf q}_1 - \hat{\bf q}_2)_i \;
        (\hat{\bf q}_1 \times 
\hat{\bf q}_2)_j + (i \leftrightarrow j) \; ,
\label{tenjet}
\end{equation}
where $\hat{\bf q}_i={\bf q}_i / |{\bf q}_i|$.
We computed the expectation values, variances etc. of the most sensitive
component $i=j=3$ of these observables. All results are shown in
figure~\ref{fig:tenjet}. In table~\ref{tab:tenjet} in
appendix~\ref{sec:numvalues}, we list the coefficients of the expectation
values (\ref{cpoddobs}) for $\OBS=T{'}_{33}^{(2)}$ (\ref{tenjet}) for
different values of the jet resolution parameter $y_{cut}$ (\ref{jade}) for
analysis 3.

%%%%%%%%%%%%%%%%%%begin%%figure%%%%%%%%%%%%%%%%%%%%%%%%%%%%%%%%%%%%%%%
\begin{figure}[ht]
  \begin{center} 
    \epsfig{bbllx=0bp,bblly=370bp,bburx=500bp,bbury=730bp,width=0.8\hsize,file=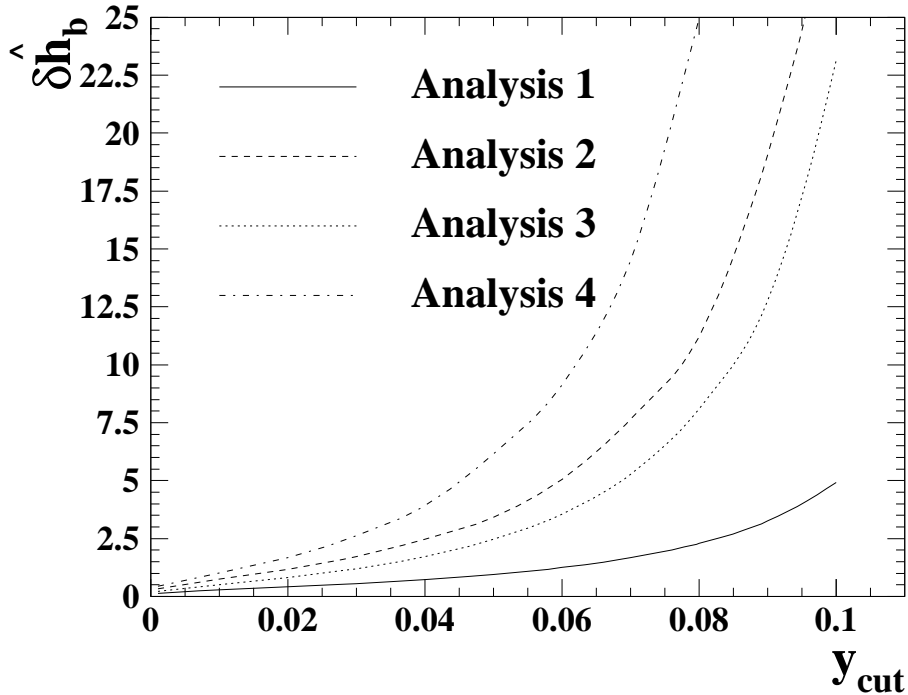}
    \parbox{0.9\textwidth}
{\caption{\it{
The error (inverse sensitivity) \dhb\ obtainable from the tensor observable
$T_{33}^{(2)}$ 
(\ref{ten}) from analysis 1 and $T{'}_{33}^{(2)}$ (\ref{tenjet}) from
analyses 2 -- 4 as function of the jet resolution parameter $y_{cut}$
(\ref{jade}) assuming (\ref{nevents}) for the number of events.
    }}\label{fig:tenjet}
}  \end{center}
\end{figure}
%%%%%%%%%%%%%%%%%%%%%%%%%%%%%%end%%figure%%%%%%%%%%%%%%%%%%%%%%%%%%%%%

%%%%%%%%%%%%%%%%%%%%%%%%%%%%%%%%%%%%%%%%%%%%%%%%%%%%%%%%%%%%%%%%%%%%%%%%
\subsubsection{Optimal observables}
%%%%%%%%%%%%%%%%%%%%%%%%%%%%%%%%%%%%%%%%%%%%%%%%%%%%%%%%%%%%%%%%%%%%%%%%

The optimal observables are given in (\ref{opti}), where $\phi$
stands for the relevant phase space variables. Note that in calculating
$S_j(\phi)$ ($j=0,1,2$) from (\ref{mezerl}) we have to sum over the
subprocesses (\ref{proc1} -- \ref{proc3}) taking into account how they
contribute to the various analyses (tables~\ref{tab:eventclassg} --
\ref{tab:eventclassq}).

In figure~\ref{fig:sens} we show the results for analysis 1 for
$y_{cut}=0.02$ in the \hvb-\hab-plane. 
Compared to the tensor and vector observables $T_{33}^{(2)}$, $V_3^{(2)}$
combined the optimal observables give only a marginal improvement now. This
is in contrast to the partonic case and shows again that a lot of
information about the CP-violating couplings is contained in the
distribution of the secondary quark and anti quark in the subprocesses
(\ref{proc2}, \ref{proc3}). This information is washed out by assuming only
knowledge of the momentum ordering of the two corresponding jets. We give
the numerical values for the elements of the coefficient matrix
(\ref{cijopt}) for the optimal observables (\ref{opti}) for different
values of the jet resolution parameter $y_{cut}$ (\ref{jade}) for analysis
1 in table~\ref{tab:cijoptjet} of appendix~\ref{sec:numvalues}.

In figure~\ref{fig:optjet} we show the inverse sensitivities \dhb\
for the 
optimal observable $O_1$ (cf. (\ref{opti})) in the analyses 2 -- 4,
as function of the jet resolution parameter. 
It is interesting to note that using the tensor observable (\ref{tenjet})
analysis 3 is superior to 2 whereas with optimal observables the reverse is
true. In table~\ref{tab:optjet} of
appendix~\ref{sec:numvalues}, we list the coefficients of the expectation
values (\ref{cpoddobs}) for $\OBS=O_1$ (\ref{opti}) for
different values of the jet resolution parameter $y_{cut}$ (\ref{jade}) for
analysis 3.

%%%%%%%%%%%%%%%%%%begin%%figure%%%%%%%%%%%%%%%%%%%%%%%%%%%%%%%%%%%%%%%
\begin{figure}[ht]
  \begin{center} 
    \epsfig{bbllx=0bp,bblly=370bp,bburx=500bp,bbury=730bp,width=0.8\hsize,file=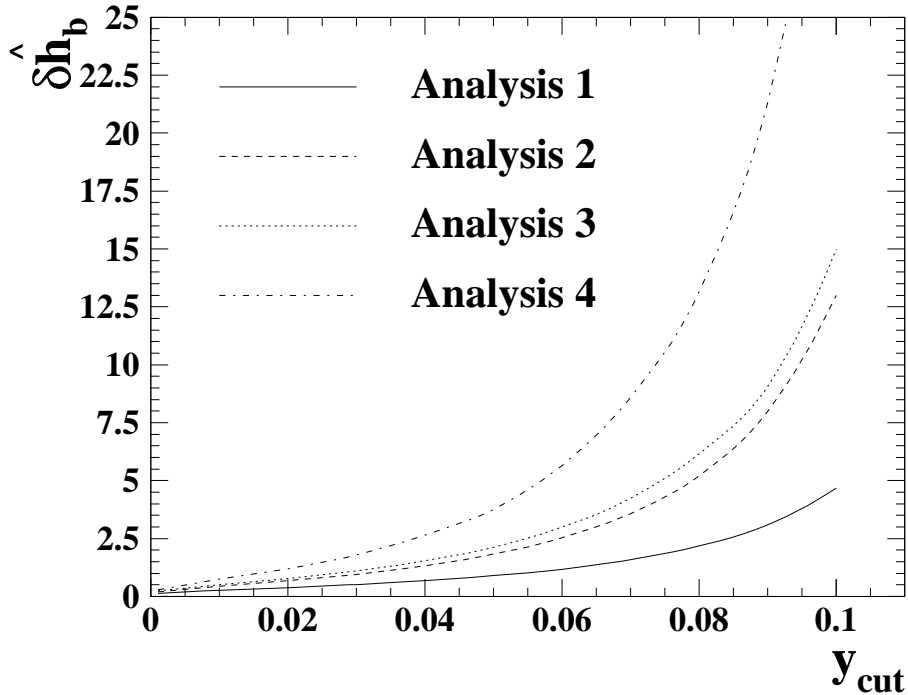}
    \parbox{0.9\textwidth}
{\caption{\it{
The error \dhb\ obtainable from the optimal observable $O_1$ (\ref{opti})
from analyses 1 -- 4 as function of the jet resolution parameter $y_{cut}$
(\ref{jade}) assuming (\ref{nevents}) for the number of events.
    }}\label{fig:optjet}
}  \end{center}
\end{figure}
%%%%%%%%%%%%%%%%%%%%%%%%%%%%%%end%%figure%%%%%%%%%%%%%%%%%%%%%%%%%%%%%

%%%%%%%%%%%%%%%%%%%%%%%%%%%%%%%%%%%%%%%%%%%%%%%%%%%%%%%%%%%%%%%%%%%%%%%%
\subsection{Comparison with the decay Z $\rightarrow$ 3 jets}
\label{sec:34jet}
%%%%%%%%%%%%%%%%%%%%%%%%%%%%%%%%%%%%%%%%%%%%%%%%%%%%%%%%%%%%%%%%%%%%%%%%

\begin{figure}[ht]
  \begin{center} 
    \epsfig{bbllx=0bp,bblly=370bp,bburx=500bp,bbury=730bp,width=0.8\hsize,file=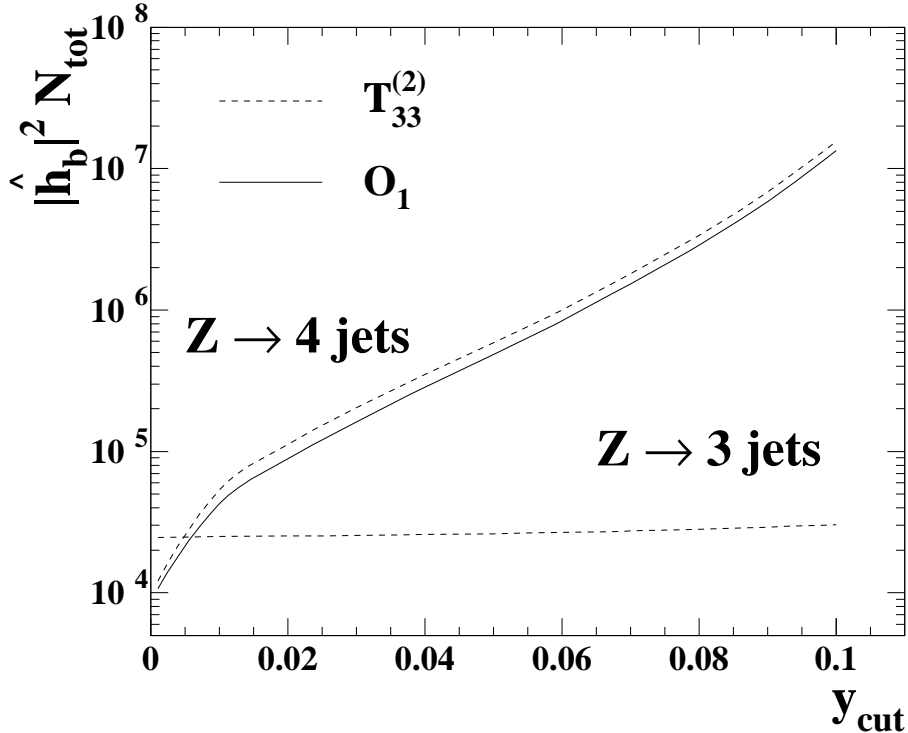}
    \parbox{0.9\textwidth}
{\caption{\it{
Comparison of the sensitivities of the best tensor observable
$T_{33}^{(2)}$ 
(\ref{ten}) and the optimal observable $O_1$ (\ref{opti}) for the \ZJ\
(analysis 1) with the best tensor 
observable $T_{33}^{(2)}$ 
(\ref{ten}) for the Z $\rightarrow$ 3 jets
analysis (cf.\cite{hab}). The results for $|\hb|^2 N_{tot}$ are shown as
function of the jet 
resolution parameter. $N_{tot}$ is the total number of Z decays
required to see an effect at the 1 s.~d. level for given $|\hb|$.
    }}\label{fig:ntota1_ten}
}  \end{center}
\end{figure}

\begin{figure}[ht]
  \begin{center} 
    \epsfig{file=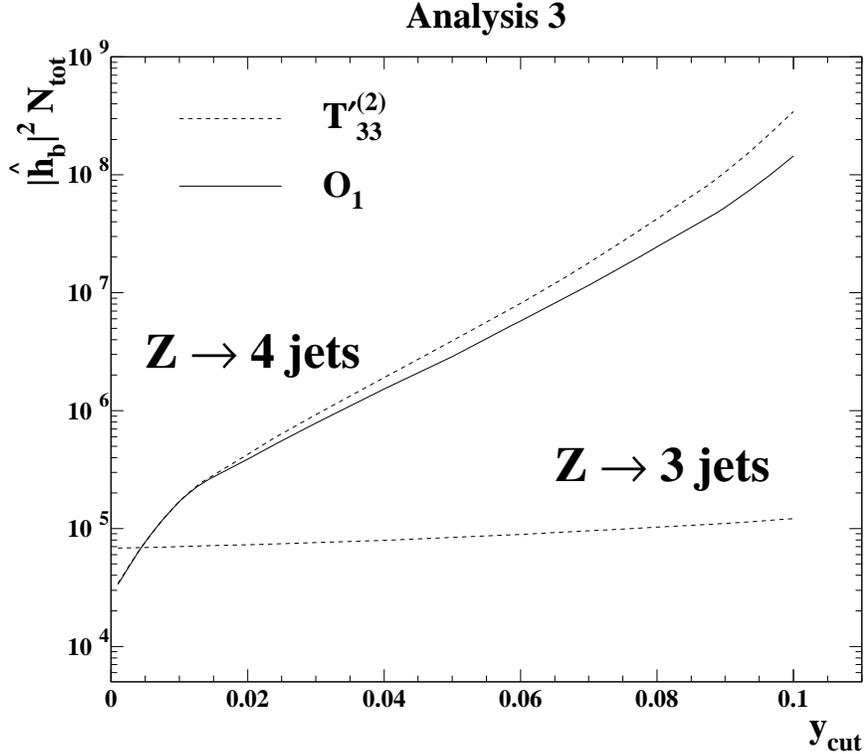,width=0.75\hsize}
    \parbox{0.9\textwidth}
{\caption{\it{
Comparison of the sensitivities of the best tensor observable
$T{'}_{33}^{(2)}$
(\ref{tenjet}) and the optimal observable $O_1$ (\ref{opti}) for the \ZJ\
(analysis 3) with the tensor 
observable $T{'}_{33}^{(2)}$ (\ref{tenjet}) for the Z $\rightarrow$ 3 jets
analysis (s. also \cite{width,hab}). The results $|\hb|^2 N_{tot}$ are shown as
function of the jet 
resolution parameter. $N_{tot}$ is the total number of Z decays
required to see an effect at the 1 s.~d. level for given $|\hb|$.  
    }}\label{fig:ntota3}
}  \end{center}
\end{figure}

Since \hbn\ is in essence only measurable with $b$ and \bb\ distinction
we concentrate on \hb\ in the following as measured with the tensor
observables and 
the optimal observable $O_1$ in analyses 1 -- 4. To compare the
sensitivities of these analyses to those from the 3 jet analyses
we calculate for each observable \OBS\ the total number of
Z events needed to measure \hb\ with a 1 s.~d. accuracy \dhb\ within the
cuts considered. 
In figures~\ref{fig:ntota1_ten}, \ref{fig:ntota3} we show
these results for analysis 1 and 3, respectively. 
Our results for the 3 jet analyses agree with the calculations
\cite{width,hab}. 
We see that the 4 jet
analyses are competitive and even better than the 3 jet analyses for small
values of the cut parameter $y_{cut}$. It should be noted, however, that
our results concern the statistical errors only. Taking into account
experimental efficiencies and systematic errors could change the situation
considerably. 

%%%%%%%%%%%%%%%%%%begin%%figure%%%%%%%%%%%%%%%%%%%%%%%%%%%%%%%%%%%%%%%
%%%%%%%%%%%%%%%%%%%%%%%%%%%%%%end%%figure%%%%%%%%%%%%%%%%%%%%%%%%%%%%%
%%%%%%%%%%%%%%%%%%begin%%figure%%%%%%%%%%%%%%%%%%%%%%%%%%%%%%%%%%%%%%%
%%%%%%%%%%%%%%%%%%%%%%%%%%%%%%end%%figure%%%%%%%%%%%%%%%%%%%%%%%%%%%%%

%%%%%%%%%%%%%%%%%%%%%%%%%%%%%%%%%%%%%%%%%%%%%%%%%%%%%%%%%%%%%%%%%%%%%%%%
%%%%%%%%%%%%%%%%%%%%%%%%%%%%%%%%%%%%%%%%%%%%%%%%%%%%%%%%%%%%%%%%%%%%%%%%
\section{Conclusions}
%%%%%%%%%%%%%%%%%%%%%%%%%%%%%%%%%%%%%%%%%%%%%%%%%%%%%%%%%%%%%%%%%%%%%%%%
%%%%%%%%%%%%%%%%%%%%%%%%%%%%%%%%%%%%%%%%%%%%%%%%%%%%%%%%%%%%%%%%%%%%%%%%

In this paper, we have presented various calculations concerning the search
for CP violation in the 4 jet decays of the Z boson with at least two of
the jets 
originating from $b$ and $\bar{b}$ quarks. We have studied a CP-violating
contact interaction with a vector and axial vector coupling \hvb, \hab\
(\ref{lcp}). Such couplings can arise at one loop level in multi-Higgs
extensions of the Standard Model \cite{higgs,bernnachhiggs}.

We found that, for reasonable values of the coupling constants, the
additional contribution of the contact interaction to the 
decay width is at most at the percent level. The decay width alone is
therefore not 
appropriate for determining the coupling constants.

We investigated tensor and vector as well as optimal
observables 
which can be used for the measurement of the
anomalous couplings.
We studied different scenarios for an experimental
analysis of the anomalous couplings: The ideal case where all the momenta
and flavours of the partons can be reconstructed from the jets and four
realistic cases where flavour information is available only for the $b$
jets.

If flavour tagging of all jets is available then, with a total number of Z
decays $N_{tot} \cong 6.4 \cdot 10^5$ and choosing a cut parameter $y_{cut}
= 0.02$ the anomalous coupling constants
\hb, \hbn\ (\ref{hat}, \ref{tilde}) can be determined with an accuracy of order
0.1 -- 0.2 at 1~s.~d. level using optimal observables (see
figs. \ref{fig:obsgg} - \ref{fig:sens}).

In the more realistic case where flavour tagging is available only for $b$ and
\bb\ jets, the coupling constant
\hb\ can be measured with an accuracy of order 0.5 -- 0.6 using the same
total number of Z 
decays. In such a measurement $b - \bb$ distinction is not necessary. Using
in particular the simple tensor observable 
$T_{33}^{(2)}$ (\ref{ten}) for the measurement, an almost optimal
sensitivity to \hb\ can be attained. 

If $b - \bb$ distinction is experimentally realizable, the coupling constant
\hbn\ can be measured with an accuracy of order 0.8 . Again we found a
simple vector observable 
$V_{3}^{(2)}$ (\ref{vec}) with an almost optimal
sensitivity to \hbn.
If $b - \bb$ distinction is experimentally not realizable
the coupling constant \hbn\ remains
essentially unconstrained 
from measurements of CP-odd observables. It can be bounded indirectly by
assuming, for instance, that its contribution to the 4 jet width does not
exceed $5 \%$. This implies then $|\hbn| \ltap\, O(1)$. 

In our theoretical investigations we assumed always $100\%$ efficiencies
and considered the statistical errors only. But the total number of Z
decays collected by the LEP and SLC experiments together is of order
$10^7$. Thus the accuracies in the determinations of \hb, \hbn\ discussed
above should indeed be within experimental reach.

Comparing 3 and 4 jet analyses we found that the sensitivity to the
anomalous coupling \hb\ was roughly constant as function of the cut
parameter \ycut\ for $\ycut<0.1$ in the 3 jet case. For the 4 jet case the
sensitivity was found to increase as \ycut\ decreases. For $\ycut\, \ltap\, 0.01$
the 4 jet sensitivity was found to exceed that from 3 jets
(figures~\ref{fig:ntota1_ten}, \ref{fig:ntota3}). Of course in an
experimental analysis one should try to make both 3 and 4 jet analyses in
order to extract the maximal possible information from the data.

For the experimental analyses, one usually has to make Monte Carlo
simulations. For this purpose one needs matrix elements including the
CP-violating interaction. These are 
available from us in the form of FORTRAN subroutines.\footnote{World
Wide Web address: {\tt http://www.thphys.uni-heidelberg.de/$\:\tilde{
    }\,$schwanen}}

To conclude: we have discussed in detail various possibilities to determine or
obtain limits on anomalous
CP-violating $Z b \bar{b} G$ and $Z b \bar{b} G G$ couplings.
As shown in \cite{higgs,bernnachhiggs} this will give valuable information
on the scalar sector in multi-Higgs extensions of the Standard Model.

\enlargethispage{\baselineskip}
%%%%%%%%%%%%%%%%%%%%%%%%%%%%%%%%%%%%%%%%%%%%%%%%%%%%%%%%%%%%%%%%%%%%%%%%
\subsection*{Acknowledgements}
%%%%%%%%%%%%%%%%%%%%%%%%%%%%%%%%%%%%%%%%%%%%%%%%%%%%%%%%%%%%%%%%%%%%%%%%
We would like to thank W.~Bernreuther, A.~Brandenburg, S.~Dhamotharan,
M. Diehl, P.~Haberl, W.~Kilian, J.~von Krogh, R.~Liebisch, P.~Overmann,
S.~Schmitt, M.~Steiert, D.~Topaj and M.~Wunsch for valuable discussions.

%%%%%%%%%%%%%%%%%%%%%%%%%%%%%%%%%%%%%%%%%%%%%%%%%%%%%%%%%%%%%%%%%%%%%%%%
%%%%%%%%%%%%%%%%%%%%%%%%%%%%%%%%%%%%%%%%%%%%%%%%%%%%%%%%%%%%%%%%%%%%%%%%
%%% Appendix
%%%%%%%%%%%%%%%%%%%%%%%%%%%%%%%%%%%%%%%%%%%%%%%%%%%%%%%%%%%%%%%%%%%%%%%%
%%%%%%%%%%%%%%%%%%%%%%%%%%%%%%%%%%%%%%%%%%%%%%%%%%%%%%%%%%%%%%%%%%%%%%%%

\begin{appendix}

%%%%%%%%%%%%%%%%%%%%%%%%%%%%%%%%%%%%%%%%%%%%%%%%%%%%%%%%%%%%%%%%%%%%%%%%
\section{Numerical Values}
\label{sec:numvalues}
%%%%%%%%%%%%%%%%%%%%%%%%%%%%%%%%%%%%%%%%%%%%%%%%%%%%%%%%%%%%%%%%%%%%%%%%

We list some numerical results for the coefficient matrices and 
covariance matrices in different studies. The statistical errors of the
numerical calculation are typically at the per cent level.

%%%%%%%%%%%%%%%%%%%begin%%table%%%%%%%%%%%%%%%%%%%%%%%%%%%%%%%%%%%%%%%
\begin{table}[H]
  \begin{center}
\begin{tabular}{|c||c|c|c|c|}
\hline
$y_{cut}$ & $c_{11}$ & $c_{12}$ & $c_{21}$ & $c_{22}$ \\
\hline \hline
0.01    & $-0.01827$  & $-2.496\cdot 10^{-5}$ & $-3.556\cdot 10^{-4}$ & $6.426\cdot 10^{-3}$\\ 
\hline                            
0.02    & $-0.02320$ & $-6.706\cdot 10^{-5}$ & $-4.504\cdot 10^{-4}$ & $8.336\cdot 10^{-3}$\\
\hline                            
0.05    & $-0.03253$ & $-1.666\cdot 10^{-4}$ & $-7.127\cdot 10^{-4}$ & $0.01248$ \\
\hline                            
\hline                            
\end{tabular}
\parbox{0.9\textwidth}{\caption{\it{
      The numerical values for the elements of the coefficient
      matrix (\ref{coeffobs}) of the
      observables $\OBS_1=T_{33}^{(2)}$ (\ref{ten}), $\OBS_2=V_{3}^{(2)}$
      (\ref{vec}) for different values of the jet resolution
      parameter $y_{cut}$ (\ref{jade}) for partons in the final state
      (section~\ref{sec:partons}).
    \label{tab:coeffobs}}}}
  \end{center}
\end{table}
%%%%%%%%%%%%%%%%%%%%%%%%%%%%%%end%%table%%%%%%%%%%%%%%%%%%%%%%%%%%%%%

%%%%%%%%%%%%%%%%%%%begin%%table%%%%%%%%%%%%%%%%%%%%%%%%%%%%%%%%%%%%%%%
\begin{table}[H]
  \begin{center}
\begin{tabular}{|c||c|c|c|}
\hline
$y_{cut}$ & $V(\OBS)_{11}$ & $V(\OBS)_{12}=V(\OBS)_{21}$ & $V(\OBS)_{22}$ \\
\hline \hline
0.01    & 0.2772     & 0.01811 & 0.1109 \\ 
\hline            
0.02    & 0.3429     & 0.02334 & 0.1427 \\
\hline            
0.05    & 0.4332   & 0.02949 & 0.1959 \\
\hline                            
\hline                            
\end{tabular}
\parbox{0.9\textwidth}{\caption{\it{
      The numerical values for the elements of the covariance
      matrix (\ref{covobs}) of the
      observables $\OBS_1=T_{33}^{(2)}$ (\ref{ten}), $\OBS_2=V_{3}^{(2)}$
      (\ref{vec}) for different values of the jet resolution
      parameter $y_{cut}$ (\ref{jade}) for partons in the final state
      (section~\ref{sec:partons}).
    \label{tab:covobs}}}}
  \end{center}
\end{table}
%%%%%%%%%%%%%%%%%%%%%%%%%%%%%%end%%table%%%%%%%%%%%%%%%%%%%%%%%%%%%%%

%%%%%%%%%%%%%%%%%%%begin%%table%%%%%%%%%%%%%%%%%%%%%%%%%%%%%%%%%%%%%%%
\begin{table}[H]
  \begin{center}
\begin{tabular}{|c||c|c|c|}
\hline
$y_{cut}$ & $c_{11}$ & $c_{12}=c_{21}$ & $c_{22}$ \\
\hline \hline
0.01    & $5.556\cdot 10^{-3}$ & $3.228\cdot 10^{-3}$ & $4.726\cdot 10^{-3}$ \\ 
\hline            
0.02    & $9.557\cdot 10^{-3}$ & $6.698\cdot 10^{-3}$ & $9.588\cdot 10^{-3}$ \\
\hline            
0.05    & 0.02274  & 0.02002  & 0.02816 \\
\hline                            
\hline                            
\end{tabular}
\parbox{0.9\textwidth}{\caption{\it{
      The numerical values for the elements of the coefficient matrix
      (\ref{cijopt}) for the optimal observables (\ref{opti}) for different
      values of the jet resolution 
      parameter $y_{cut}$ (\ref{jade}) for partons in the final state
      (section~\ref{sec:partons}).
    \label{tab:cijopt}}}}
  \end{center}
\end{table}
%%%%%%%%%%%%%%%%%%%%%%%%%%%%%%end%%table%%%%%%%%%%%%%%%%%%%%%%%%%%%%%

%%%%%%%%%%%%%%%%%%%begin%%table%%%%%%%%%%%%%%%%%%%%%%%%%%%%%%%%%%%%%%%
\begin{table}[H]
  \begin{center}
\begin{tabular}{|c||c|}
\hline
$y_{cut}$ & $c_1$ \\
\hline \hline
0.01    & $-0.01020$ \\ 
\hline                
0.02    & $-0.01256$ \\
\hline               
0.05    & $-0.01502$ \\
\hline                            
\hline                            
\end{tabular}
\parbox{0.9\textwidth}{\caption{\it{
      The numerical values the coefficients of the expectation
      values (\ref{cpoddobs}) for $\OBS=T{'}_{33}^{(2)}$ (\ref{tenjet}) for
      different values of the jet resolution parameter $y_{cut}$
      (\ref{jade}) for analysis 3 of section~\ref{sec:jets}.
    \label{tab:tenjet}}}}
  \end{center}
\end{table}
%%%%%%%%%%%%%%%%%%%%%%%%%%%%%%end%%table%%%%%%%%%%%%%%%%%%%%%%%%%%%%%

%%%%%%%%%%%%%%%%%%%begin%%table%%%%%%%%%%%%%%%%%%%%%%%%%%%%%%%%%%%%%%%
\begin{table}[H]
  \begin{center}
\begin{tabular}{|c||c|c|c|}
\hline
$y_{cut}$ & $c_{11}$ & $c_{12}=c_{21}$ & $c_{22}$ \\
\hline \hline
0.01    & $1.514\cdot 10^{-3}$ & $1.041\cdot 10^{-4}$ & $4.825\cdot 10^{-4}$\\ 
\hline            
0.02    & $1.981\cdot 10^{-3}$ & $1.651\cdot 10^{-4}$ & $6.396\cdot 10^{-4}$\\
\hline           
0.05    & $2.956\cdot 10^{-3}$ & $3.042\cdot 10^{-4}$ & $9.907\cdot 10^{-4}$ \\
\hline                            
\hline                            
\end{tabular}
\parbox{0.9\textwidth}{\caption{\it{
      The numerical values for the elements of the coefficient
      matrix (\ref{cijopt}) for the optimal observables (\ref{opti}) for
      different values of the jet resolution parameter $y_{cut}$
      (\ref{jade}) for analysis 1 of section~\ref{sec:jets}.
    \label{tab:cijoptjet}}}}
  \end{center}
\end{table}
%%%%%%%%%%%%%%%%%%%%%%%%%%%%%%end%%table%%%%%%%%%%%%%%%%%%%%%%%%%%%%%

%%%%%%%%%%%%%%%%%%%begin%%table%%%%%%%%%%%%%%%%%%%%%%%%%%%%%%%%%%%%%%%
\begin{table}[H]
  \begin{center}
\begin{tabular}{|c||c|}
\hline
$y_{cut}$ & $c_1$ \\
\hline \hline
0.01    & $5.535\cdot 10^{-4}$\\ 
\hline      
0.02    & $7.124\cdot 10^{-4}$\\
\hline     
0.05    & $8.627\cdot 10^{-4}$\\
\hline                            
\hline                            
\end{tabular}
\parbox{0.9\textwidth}{\caption{\it{
      The numerical values for the coefficients of the expectation
      values (\ref{cpoddobs}) for the optimal observable $\OBS=O_1$
      (\ref{opti}) for
      different values of the jet resolution parameter $y_{cut}$
      (\ref{jade}) for analysis 3 of section~\ref{sec:jets}.
    \label{tab:optjet}}}}
  \end{center}
\end{table}
%%%%%%%%%%%%%%%%%%%%%%%%%%%%%%end%%table%%%%%%%%%%%%%%%%%%%%%%%%%%%%%

%%%%%%%%%%%%%%%%%%%%%%%%%%%%%%%%%%%%%%%%%%%%%%%%%%%%%%%%%%%%%%%%%%%%%%%%
\section{Eventclasses}
\label{sec:eventclass}
%%%%%%%%%%%%%%%%%%%%%%%%%%%%%%%%%%%%%%%%%%%%%%%%%%%%%%%%%%%%%%%%%%%%%%%%

Here we explain which classes of events contribute to the four different
analyses as defined in chapter \ref{sec:jets}. First, we compare
the partonic phase space with the jet phase space.

\renewcommand{\arraystretch}{1.4}
%%%%%%%%%%%%%%%%%%%begin%%table%%%%%%%%%%%%%%%%%%%%%%%%%%%%%%%%%%%%%%%
\begin{table}[H]
  \begin{center}
\begin{tabular}{|c|c|}
\hline
Process & Phase space restriction\\
\hline \hline
\EZG\ & $|{\bf k}_1| \geq |{\bf k}_2|$ \\
\hline                            
\EZB\ & $|{\bf k}_+| \geq |{\bf q}_+|$, $|{\bf k}_-| \geq |{\bf q}_-| $ \\
\hline                            
\EZQ\ & --- \\
\hline                            
\end{tabular}
\parbox{0.9\textwidth}{\caption{\it{
      The restrictions on the partonic phase space for the different
      processes due to identical particles in the final state.
    \label{tab:partphasesp}}}}
  \end{center}
\end{table}
%%%%%%%%%%%%%%%%%%%%%%%%%%%%%%end%%table%%%%%%%%%%%%%%%%%%%%%%%%%%%%%

%%%%%%%%%%%%%%%%%%%begin%%table%%%%%%%%%%%%%%%%%%%%%%%%%%%%%%%%%%%%%%%
\begin{table}[H]
  \begin{center}
\begin{tabular}{|c|c|}
\hline
Analysis & Phase space restriction\\
\hline \hline
1 & $|{\bf q}_3| \geq |{\bf q}_4|$ \\
\hline
2, 3, 4 & $|{\bf q}_1| \geq |{\bf q}_2| \geq |{\bf q}_3| \geq |{\bf q}_4|$ \\
\hline                            
\end{tabular}
\parbox{0.9\textwidth}{\caption{\it{
      The restrictions on the jet phase space in the analyses 1 -- 4. 
    \label{tab:jetphasesp}}}}
  \end{center}
\end{table}
%%%%%%%%%%%%%%%%%%%%%%%%%%%%%%end%%table%%%%%%%%%%%%%%%%%%%%%%%%%%%%%

In Tables~\ref{tab:partphasesp},~\ref{tab:jetphasesp} we list
the restrictions on the phase space for the partonic processes
(\ref{proc1} -- \ref{proc3}) and for the jets in
the analyses 1 -- 4 as defined in chapter~\ref{sec:jets}.

In tables \ref{tab:eventclassg} -- \ref{tab:eventclassq} we list all
possibilities how the 4 partons of the reactions (\ref{proc1} --
\ref{proc3}) can give 4 jets with the ordering criteria of the analyses 1
-- 4. The full points indicate that an event class satisfies the respective
selection criterion.

%\newpage
%%%%%%%%%%%%%%%%%%begin%%table%%%%%%%%%%%%%%%%%%%%%%%%%%%%%%%%%%%%%%%
\begin{table}[H]
  \begin{center}
%\underline{\EZG} \\
%\vspace{0.5cm}
\begin{tabular}{|c|c|c|c||c|}
\hline
\ \ \ \bb\ jet \ \ \  & \ \ \ $b$ jet \ \ \ & \ \ \ jet 3 \ \ \  & \ \ \
jet 4 \ \ \ & Analysis 1 \\
\hline \hline
$\bb(k_+)$ & $b(k_-)$ & $G(k_1)$ & $G(k_2)$ & $\bullet$ \\
\hline
\end{tabular}
%\parbox{0.9\textwidth}{\caption{\it{
%     The only possibility for the partons in \EZG\ to fulfill the selection
%     criterion of Analysis 1.
%    \label{tab:eventclassga1}}}}
%  \end{center}
%\end{table}
%%%%%%%%%%%%%%%%%%%%%%%%%%%%%end%%table%%%%%%%%%%%%%%%%%%%%%%%%%%%%%
\vspace{0.5cm}

%%%%%%%%%%%%%%%%%%%begin%%table%%%%%%%%%%%%%%%%%%%%%%%%%%%%%%%%%%%%%%%
%\begin{table}[htb]
%  \begin{center}
\begin{tabular}{|c|c|c|c||c|c|c|}
\hline
\ \ \ jet 1 \ \ \ & \ \ \ jet 2 \ \ \ & \ \ \ jet 3 \ \ \ & \ \ \ jet 4  \ \ \ & Analysis 2 & Analysis 3 & Analysis 4 \\
\hline \hline
$\bar{b}(k_+)$ & $b(k_-)$ & $G(k_1)$ & $G(k_2)$ & $\bullet$ & $\bullet$ & $\bullet$ \\ 
\hline                            
$\bar{b}(k_+)$ & $G(k_1)$ & $b(k_-)$ & $G(k_2)$ & $\bullet$ &  & $\bullet$ \\
\hline                          
$\bar{b}(k_+)$ & $G(k_1)$ & $G(k_2)$ & $b(k_-)$ & $\bullet$ &  & $\bullet$ \\
\hline                          
$b(k_-)$ & $\bar{b}(k_+)$ & $G(k_1)$ & $G(k_2)$ & $\bullet$ & $\bullet$ & $\bullet$ \\
\hline                          
$b(k_-)$ & $G(k_1)$ & $\bar{b}(k_+)$ & $G(k_2)$ & $\bullet$ &  & $\bullet$ \\
\hline                          
$b(k_-)$ & $G(k_1)$ & $G(k_2)$ & $\bar{b}(k_+)$ & $\bullet$ &  & $\bullet$ \\
\hline                          
$G(k_1)$ & $\bar{b}(k_+)$ & $b(k_-)$ & $G(k_2)$ &  & $\bullet$ & $\bullet$ \\
\hline                          
$G(k_1)$ & $\bar{b}(k_+)$ & $G(k_2)$ & $b(k_-)$ &  & $\bullet$ & $\bullet$ \\
\hline                          
$G(k_1)$ & $b(k_-)$ & $\bar{b}(k_+)$ & $G(k_2)$ &  & $\bullet$ & $\bullet$ \\
\hline                          
$G(k_1)$ & $b(k_-)$ & $G(k_2)$ & $\bar{b}(k_+)$ &  & $\bullet$ & $\bullet$ \\
\hline                          
$G(k_1)$ & $G(k_2)$ & $\bar{b}(k_+)$ & $b(k_-)$ &  &  & $\bullet$ \\
\hline                          
$G(k_1)$ & $G(k_2)$ & $b(k_-)$ & $\bar{b}(k_+)$ &  &  & $\bullet$ \\
\hline
\end{tabular}
\parbox{0.9\textwidth}{\caption{\it{
      The only possibility for the partons in \EZG\ to fulfill the selection
      criterion of analysis 1 and
      the 12 possibilities for them to give 4 momentum ordered
      jets. It is indicated by dots which event class contributes to
      which of the analyses.
      \label{tab:eventclassg}}}}
   \end{center}
\end{table}
%%%%%%%%%%%%%%%%%%%%%%%%%%%%%%%end%%table%%%%%%%%%%%%%%%%%%%%%%%%%%%%%
\newpage

%%%%%%%%%%%%%%%%%%%begin%%table%%%%%%%%%%%%%%%%%%%%%%%%%%%%%%%%%%%%%%%
\begin{table}[H]
  \begin{center}
\begin{tabular}{|c|c|c|c||c|}
\hline
\ \ \ \bb\ jet \ \ \ & \ \ \ $b$ jet \ \ \ & \ \ \ jet 3 \ \ \ & \ \ \ jet 4 \ \ \ & Analysis 1 \\
\hline \hline
$\bb(k_+)$ & $b(k_-)$ & $b(q_-)$ & $\bb(q_+)$ & $\bullet$ \\
\hline
$\bb(k_+)$ & $b(k_-)$ & $\bb(q_+)$ & $b(q_-)$ & $\bullet$ \\
\hline
$\bb(q_+)$ & $b(k_-)$ & $b(q_-)$ & $\bb(k_+)$ & $\bullet$ \\
\hline
$\bb(q_+)$ & $b(k_-)$ & $\bb(k_+)$ & $b(q_-)$ & $\bullet$ \\
\hline
$\bb(k_+)$ & $b(q_-)$ & $b(k_-)$ & $\bb(q_+)$ & $\bullet$ \\
\hline
$\bb(k_+)$ & $b(q_-)$ & $\bb(q_+)$ & $b(k_-)$ & $\bullet$ \\
\hline
$\bb(q_+)$ & $b(q_-)$ & $b(k_-)$ & $\bb(k_+)$ & $\bullet$ \\
\hline
$\bb(q_+)$ & $b(q_-)$ & $\bb(k_+)$ & $b(k_-)$ & $\bullet$ \\
\hline
\end{tabular}
%\parbox{0.9\textwidth}{\caption{\it{
%     The 8 possibilities for the partons in \EZB\ to fulfill the selection
%     criterion of analysis 1.
%    \label{tab:eventclassba1}}}}
%  \end{center}
%\end{table}
%%%%%%%%%%%%%%%%%%%%%%%%%%%%%%end%%table%%%%%%%%%%%%%%%%%%%%%%%%%%%%%
\vspace{0.5cm}

%%%%%%%%%%%%%%%%%%begin%%table%%%%%%%%%%%%%%%%%%%%%%%%%%%%%%%%%%%%%%%
%\begin{table}[H]
%  \begin{center}
\begin{tabular}{|c|c|c|c||c|c|c|}
\hline
\ \ \ jet 1 \ \ \ & \ \ \ jet 2 \ \ \ & \ \ \ jet 3 \ \ \ & \ \ \ jet 4 \ \ \ & Analysis 2 & Analysis 3 & Analysis 4 \\
\hline \hline
$\bar{b}(k_+)$ & $b(k_-)$ & $b(q_-)$ & $\bar{b}(q_+)$ & $\bullet$ & $\bullet$ & $\bullet$ \\ 
\hline                            
$\bar{b}(k_+)$ & $b(k_-)$ & $\bar{b}(q_+)$ & $b(q_-)$ & $\bullet$ & $\bullet$ & $\bullet$ \\
\hline                            
$\bar{b}(k_+)$ & $\bar{b}(q_+)$ & $b(k_-)$ & $b(q_-)$ & $\bullet$ & $\bullet$ & $\bullet$ \\
\hline                            
$b(k_-)$ & $\bar{b}(k_+)$ & $b(q_-)$ & $\bar{b}(q_+)$ & $\bullet$ & $\bullet$ & $\bullet$ \\
\hline                            
$b(k_-)$ & $\bar{b}(k_+)$ & $\bar{b}(q_+)$ & $b(q_-)$ & $\bullet$ & $\bullet$ & $\bullet$ \\
\hline                            
$b(k_-)$ & $b(q_-)$ & $\bar{b}(k_+)$ & $\bar{b}(q_+)$ & $\bullet$ & $\bullet$ & $\bullet$ \\
\hline                            
\end{tabular}
\parbox{0.9\textwidth}{\caption{\it{
     The 8 possibilities for the partons in \EZB\ to fulfill the selection
     criterion of analysis 1 and
     the 6 possibilities for them to give 4 momentum ordered
     jets. It is indicated by dots which event class contributes to which
     of the analyses.
    \label{tab:eventclassb}}}}
  \end{center}
\end{table}
%%%%%%%%%%%%%%%%%%%%%%%%%%%%%%end%%table%%%%%%%%%%%%%%%%%%%%%%%%%%%%%
\newpage

%%%%%%%%%%%%%%%%%%%begin%%table%%%%%%%%%%%%%%%%%%%%%%%%%%%%%%%%%%%%%%%
\begin{table}[H]
  \begin{center}
\begin{tabular}{|c|c|c|c||c|}
\hline
\ \ \ \bb\ jet \ \ \ & \ \ \ $b$ jet \ \ \ & \ \ \ jet 3 \ \ \ & \ \ \ jet 4 \ \ \ & Analysis 1 \\
\hline \hline
$\bb(k_+)$ & $b(k_-)$ & $q(q_-)$ & $\bq(q_+)$ & $\bullet$ \\
\hline
$\bb(k_+)$ & $b(k_-)$ & $\bq(q_+)$ & $q(q_-)$ & $\bullet$ \\
\hline
\end{tabular}
%\parbox{0.9\textwidth}{\caption{\it{
%     The 2 possibilities for the partons in \EZQ\ to fulfill the selection
%     criterion of analysis 1.
%    \label{tab:eventclassqa1}}}}
%  \end{center}
%\end{table}
%%%%%%%%%%%%%%%%%%%%%%%%%%%%%%%end%%table%%%%%%%%%%%%%%%%%%%%%%%%%%%%%
\vspace{0.5cm}

\thispagestyle{empty}
%%%%%%%%%%%%%%%%%%%begin%%table%%%%%%%%%%%%%%%%%%%%%%%%%%%%%%%%%%%%%%%
%\begin{table}[H]
%  \begin{center}
\begin{tabular}{|c|c|c|c||c|c|c|}
\hline
\ \ \ jet 1 \ \ \ & \ \ \ jet 2 \ \ \ & \ \ \ jet 3 \ \ \ & \ \ \ jet 4 \ \ \ & Analysis 2 & Analysis 3 & Analysis 4 \\
\hline \hline
$\bar{b}(k_+)$ & $b(k_-)$ & $q(q_-)$ & $\bq(q_+)$ & $\bullet$ & $\bullet$ & $\bullet$ \\ 
\hline                            
$\bar{b}(k_+)$ & $b(k_-)$ & $\bq(q_+)$ & $q(q_-)$ & $\bullet$ & $\bullet$ & $\bullet$ \\ 
\hline                            
$\bar{b}(k_+)$ & $q(q_-)$ & $b(k_-)$ & $\bq(q_+)$ & $\bullet$ &  & $\bullet$ \\
\hline                            
$\bar{b}(k_+)$ & $\bq(q_+)$ & $b(k_-)$ & $q(q_-)$ & $\bullet$ &  & $\bullet$ \\
\hline                            
$\bar{b}(k_+)$ & $q(q_-)$ & $\bq(q_+)$ & $b(k_-)$ & $\bullet$ &  & $\bullet$ \\
\hline                            
$\bar{b}(k_+)$ & $\bq(q_+)$ & $q(q_-)$ & $b(k_-)$ & $\bullet$ &  & $\bullet$ \\
\hline                            
$b(k_-)$ & $\bar{b}(k_+)$ & $q(q_-)$ & $\bq(q_+)$ & $\bullet$ & $\bullet$ & $\bullet$ \\
\hline                            
$b(k_-)$ & $\bar{b}(k_+)$ & $\bq(q_+)$ & $q(q_-)$ & $\bullet$ & $\bullet$ & $\bullet$ \\
\hline                            
$b(k_-)$ & $q(q_-)$ & $\bar{b}(k_+)$ & $\bq(q_+)$ & $\bullet$ &  & $\bullet$ \\
\hline                            
$b(k_-)$ & $\bq(q_+)$ & $\bar{b}(k_+)$ & $q(q_-)$ & $\bullet$ &  & $\bullet$ \\
\hline                            
$b(k_-)$ & $q(q_-)$ & $\bq(q_+)$ & $\bar{b}(k_+)$ & $\bullet$ &  & $\bullet$ \\
\hline                            
$b(k_-)$ & $\bq(q_+)$ & $q(q_-)$ & $\bar{b}(k_+)$ & $\bullet$ &  & $\bullet$ \\
\hline                            
$q(q_-)$ & $\bar{b}(k_+)$ & $b(k_-)$ & $\bq(q_+)$ &  & $\bullet$ & $\bullet$ \\
\hline                            
$\bq(q_+)$ & $\bar{b}(k_+)$ & $b(k_-)$ & $q(q_-)$ &  & $\bullet$ & $\bullet$ \\
\hline                            
$q(q_-)$ & $\bar{b}(k_+)$ & $\bq(q_+)$ & $b(k_-)$ &  & $\bullet$ & $\bullet$ \\
\hline                            
$\bq(q_+)$ & $\bar{b}(k_+)$ & $q(q_-)$ & $b(k_-)$ &  & $\bullet$ & $\bullet$ \\
\hline                            
$q(q_-)$ & $b(k_-)$ & $\bar{b}(k_+)$ & $\bq(q_+)$ &  & $\bullet$ & $\bullet$ \\
\hline                            
$\bq(q_+)$ & $b(k_-)$ & $\bar{b}(k_+)$ & $q(q_-)$ &  & $\bullet$ & $\bullet$ \\
\hline                            
$q(q_-)$ & $b(k_-)$ & $\bq(q_+)$ & $\bar{b}(k_+)$ &  & $\bullet$ & $\bullet$ \\
\hline                            
$\bq(q_+)$ & $b(k_-)$ & $q(q_-)$ & $\bar{b}(k_+)$ &  & $\bullet$ & $\bullet$ \\
\hline                            
$q(q_-)$ & $\bq(q_+)$ & $\bar{b}(k_+)$ & $b(k_-)$ &  &  & $\bullet$ \\
\hline                            
$\bq(q_+)$ & $q(q_-)$ & $\bar{b}(k_+)$ & $b(k_-)$ &  &  & $\bullet$ \\
\hline                            
$q(q_-)$ & $\bq(q_+)$ & $b(k_-)$ & $\bar{b}(k_+)$ &  &  & $\bullet$ \\
\hline                            
$\bq(q_+)$ & $q(q_-)$ & $b(k_-)$ & $\bar{b}(k_+)$ &  &  & $\bullet$ \\
\hline
\end{tabular}
\parbox{0.9\textwidth}{\caption{\it{
     The 2 possibilities for the partons in \EZQ\ to fulfill the selection
     criterion of analysis 1 and
     the 24 possibilities for them to give 4 momentum ordered
     jets. It is indicated by dots which event class contributes to which
     of the analyses.
    \label{tab:eventclassq}}}}
  \end{center}
\end{table}
%%%%%%%%%%%%%%%%%%%%%%%%%%%%%%end%%table%%%%%%%%%%%%%%%%%%%%%%%%%%%%%
\newpage

\end{appendix}

\newpage

%%%%%%%%%%%%%%%%%%%%%%%%%%%%%%%%%%%%%%%%%%%%%%%%%%%%%%%%%%%%%%%%%%%%%%%%
%%% References
%%%%%%%%%%%%%%%%%%%%%%%%%%%%%%%%%%%%%%%%%%%%%%%%%%%%%%%%%%%%%%%%%%%%%%%%
%\baselineskip15pt
%\begin{thebibliography}{19}

\thispagestyle{plain}
\pagestyle{empty}
\end{document}